\documentclass[pdflatex,sn-mathphys-num]{sn-jnl}

\usepackage{graphicx}%
\usepackage{subcaption}%
\usepackage{multirow}%
\usepackage{amsmath,amssymb,amsfonts}%
\usepackage{amsthm}%
\usepackage{mathrsfs}%
\usepackage[title]{appendix}%
\usepackage{xcolor}%
\usepackage{textcomp}%
\usepackage{manyfoot}%
\usepackage{booktabs}%
\usepackage{algorithm}%
\usepackage{algorithmicx}%
\usepackage{algpseudocode}%
\usepackage{listings}%
\usepackage{epstopdf}
\usepackage{comment}

\begin{document}

\title{Scaling Dependencies in Irradiation-Driven Molecular Dynamics Simulations: Case Study of W(CO)$_6$ Fragmentation}


\author[1]{Soumyo Kheto}
\email{24dr0181@iitism.ac.in}

\author[2]{Alexey Verkhovtsev}
\email{verkhovtsev@mbnresearch.com}

\author[1]{Bobby Antony}
\email{bobby@iitism.ac.in}

\author[2]{Andrey V. Solov'yov}
\email{solovyov@mbnresearch.com}

\affil[1]{Department of Physics, IIT(ISM) Dhanbad, 826004 Dhanbad, Jharkhand, India}

\affil[2]{MBN Research Center, Altenhöferallee 3, 60438 Frankfurt am Main, Germany}


\abstract{Irradiation-driven fragmentation and chemical transformations of organometallic molecules play a central role in nanofabrication techniques based on the use of focused charged-particle beams. In this paper, the electron irradiation-induced fragmentation dynamics of W(CO)$_6$, a commonly used precursor for focused electron beam-induced deposition (FEBID), is investigated using the irradiation-driven molecular dynamics (IDMD) method. Simulations are performed for gas-phase systems with different precursor densities and under different irradiation conditions. The results reveal progressive fragmentation of W(CO)$_6$ molecules into W(CO)$_n$ species, accompanied by the formation of W-rich molecular clusters. The evolution of fragment abundances shows a strong dependence on both precursor density and electron fluence. Higher densities and larger fluences result in more extensive fragmentation and promote the aggregation of tungsten atoms into small metal clusters. Under certain irradiation conditions, the studied molecular systems evolve towards a steady state characterised by slightly varying fragment abundances. 
The obtained scaling relations between irradiation parameters and fragment distributions provide guidance for selecting simulation parameters in IDMD simulations of the FEBID process, ensuring a quantitative description of precursor fragmentation dynamics.}

\maketitle

\section{Introduction}

\linespread{1.2}

Irradiation-driven chemistry processes induced by the interaction of various types of radiation with molecular systems are exploited in a variety of modern and emerging technologies.
One particularly promising application is the fabrication and processing of materials at the nanoscale using focused charged-particle beams \cite{Utke2008, Utke_book_2012, DeTeresa-book2020, Barth_2025_AdvFunctMater_review}.

Electron irradiation-induced chemistry plays a central role in focused electron beam-induced deposition (FEBID), a technique that enables the fabrication of complex two- and three-dimensional structures with nanometre resolution \cite{Huth_2012_BJN.3.597, Utke2022_CoordChemRev, Huth_2021_JAP_review}. 
In FEBID, precursor molecules (typically organometallic) are exposed to an electron beam, which induces molecular fragmentation and subsequent chemical transformations, resulting in the formation and growth of metal-containing deposits. This approach enables direct-write 3D nanoprinting of structures with spatial resolutions on the order of several tens of nanometres \cite{Winkler_2019_JAP_review, Huth_2021_JAP_review, Reisecker_FEBID_review_2024, Barth_2025_AdvFunctMater_review}.

Despite significant progress made in this research area over the past two decades \cite{Utke2008, Utke_book_2012, DeTeresa-book2020, Barth_2025_AdvFunctMater_review}, the fundamental radiation-induced physicochemical processes governing the formation and growth of nanostructures under focused particle-beam irradiation remain not fully understood and are the subject of active ongoing research \cite{Winkler_2019_JAP_review, Huth_2021_JAP_review, Reisecker_FEBID_review_2024, Barth_2025_AdvFunctMater_review, DySoN_book_Springer_2022, Roadmap_ChemRev2024, SD_FEBID_arXiv}. 
In recent years, considerable effort has been devoted to elucidating the elementary processes  that lead to electron-induced cleavage of metal--ligand bonds in FEBID precursors; see, for example, review papers \cite{Thorman2015, Barth2020_JMaterChemC, Utke2022_CoordChemRev} and references therein. To date, however, most available information on electron-irradiation-induced processes involving FEBID precursors has been obtained from experimental studies.


Atomistic insight into irradiation-driven chemical processes in molecular systems, such as bond cleavage and subsequent chemical reactivity, can be obtained through computational modelling.
A rigorous quantum-mechanical description of these processes, e.g. within time-dependent density functional theory (TDDFT) \cite{Ullrich_TDDFT_book} or quantum molecular dynamics (MD) approaches \cite{Marx_Hutter_AIMD_2009}, is feasible only for relatively small systems containing up to a few hundred atoms and is typically limited to timescales of several tens to hundreds of femtoseconds \cite{Springer_Handbook_CompChem}.

Radiation- and collision-induced fragmentation of molecular systems, as well as post-irradiation dynamics occurring on much longer time scales (nanoseconds and beyond), have been successfully investigated using classical reactive MD \cite{Sushko2016_rCHARMM} and Irradiation-Driven MD (IDMD) \cite{Sushko2016_IDMD} methods implemented in the advanced software package MBN Explorer \cite{Solovyov_2012_JCC_MBNExplorer}. 
These methods enable the incorporation of random, fast and local quantum transformations occurring in molecular systems due to chemical reactions or irradiation-induced quantum processes (e.g. bond breakage and formation) into the classical MD framework \cite{Sushko2016_IDMD}. This makes it possible to simulate chemical and radiation-driven transformations in molecular and condensed matter systems on temporal and spatial scales that are inaccessible to \textit{ab initio} methods \cite{MBNbook_Springer_2017, DySoN_book_Springer_2022, Roadmap_ChemRev2024}.

Within the IDMD framework \cite{Sushko2016_IDMD}, various quantum processes occurring in an irradiated system (such as bond dissociation via ionisation or electron attachment, charge transfer, etc.) are incorporated stochastically into the classical MD framework. The probabilities of these processes are determined from the corresponding cross sections. Major transformations of irradiated molecular systems, including changes in molecular topology, modifications of interatomic interactions, and redistribution of atomic partial charges, are simulated using reactive rCHARMM force fields \cite{Sushko2016_rCHARMM} with MBN Explorer \cite{Solovyov_2012_JCC_MBNExplorer}.


The IDMD method has been successfully used to provide detailed atomistic characterisation of nanostructures grown by FEBID \cite{Sushko2016_IDMD, DeVera2020, Prosvetov2021_BJN, Prosvetov2022_PCCP, Prosvetov2023_EPJD}, as well as to describe radiation-driven chemistry involving several organometallic FEBID precursors \cite{DeVera2019, Andreides_2023_JPCA_FeCO5, Lyshchuk_2025_JPCA_MeCpPtMe3}. The IDMD simulations of the FEBID process enabled prediction of key properties of the deposits, including their geometrical characteristics (lateral size, height, and volume), morphology, growth rate, and chemical composition.

In FEBID experiments, precursor molecules are typically exposed to the electron beam for dwell times ranging from hundreds of microseconds to milliseconds \cite{Huth_2012_BJN.3.597}. Such timescales are far beyond those accessible in all-atom MD simulations. To overcome this obstacle, in the previous IDMD simulations of the FEBID process \cite{Sushko2016_IDMD, Prosvetov2021_BJN, DeVera2020, Prosvetov2022_PCCP, Prosvetov2023_EPJD}, the primary electron flux was rescaled so that the cumulative number of electrons interacting with the system during the simulated dwell time (i.e. integral electron fluence) reproduced the experimental value \cite{Sushko2016_IDMD}.
These simulations were performed using a constant dwell time of 10~ns and the electron beam current of one to several microamperes. Simulating multiple consecutive cycles of precursor deposition and irradiation enabled studying the growth of FEBID deposits over timescales of several hundred nanoseconds \cite{Sushko2016_IDMD, Prosvetov2021_BJN, DeVera2020, Prosvetov2022_PCCP, Prosvetov2023_EPJD}.

An important question that arises in this context is to what extent the radiation-induced dynamics of precursor molecules depends on the dwell time and electron beam current used in simulations, even when the total electron fluence is kept constant. This question is systematically addressed in the present study, which focuses on a computational investigation of electron irradiation-driven fragmentation dynamics and chemistry of W(CO)$_6$ precursors using the IDMD method. 
The analysis considers electron-induced chemistry of W(CO)$_6$ molecular ensembles in the gas phase, their fragmentation, and the subsequent formation of metal-containing molecular clusters under different irradiation conditions. Specifically, we investigate how precursor fragmentation depends on electron beam current and irradiation time for a given total electron fluence, as well as on precursor density. 

The performed analysis enables us to establish relationships between irradiation conditions, precursor density, and the resulting fragmentation patterns, i.e. the relative abundance of different molecular species. In particular, we determine branching ratios for different fragments and analyse conditions for the formation of metal-containing molecular aggregates.

Our earlier studies of radiation-induced fragmentation of FEBID precursors \cite{DeVera2019, Andreides_2023_JPCA_FeCO5, Lyshchuk_2025_JPCA_MeCpPtMe3} considered the fragmentation of a single molecule. The novel aspect of this study is the analysis of fragmentation dynamics for an ensemble of molecules.

The systems and processes considered in this study are also relevant to recent experimental investigations on W(CO)$_6$ fragmentation using the focused electron beam induced mass spectrometry (FEBiMS) method \cite{FEBIMS}, which enables real-time monitoring of electron-induced chemical processes through \textit{in situ} mass detection of charged fragments generated during FEBID. Recent investigations of W(CO)$_6$ fragmentation using FEBiMS \cite{FEBIMS} have shown that the primary fragmentation channel involves the detachment of CO ligands, accompanied by the formation of positively and negatively charged fragments, W(CO)$_n^{+}$ and W(CO)$_n^{-}$ ($n = 0-6$), consistent with earlier cross-beam experiments on gas-phase W(CO)$_6$ systems \cite{WNOROSKI-R, WNOROWSKI-IJMS}. The mass spectra recorded during FEBID with W(CO)$_6$ precursors \cite{FEBIMS} included both volatile, metal-free fragments and heavier species, including W(CO)$_{6-n}^+$ ($n=1-5$) and bare W$^+$ ions. These observations have stimulated discussion regarding the origin of fragment signals in FEBiMS mass spectra, which may arise either from desorption of ionised fragments from the substrate or from ionisation events involving precursor molecules and their fragments in the gas phase \cite{Jureddy_PCCP_2025}.

The results of the present study provide useful information for selecting appropriate dwell times in future IDMD simulations of the FEBID process, while providing a quantitative description of precursor fragmentation dynamics. In addition, the presented results may stimulate further experimental investigations of electron-induced fragmentation dynamics of FEBID precursors using advanced techniques such as FEBiMS.

\section{Methodology}
\label{sec:Methodology}

The IDMD simulations were performed using MBN Explorer \cite{Solovyov_2012_JCC_MBNExplorer}, a software package for multiscale simulations of the structure and dynamics of complex Meso-Bio-Nano (MBN) systems \cite{MBNbook_Springer_2017}. The MBN Studio toolkit \cite{Sushko_2019_MBNStudio} was used to create the systems, prepare the necessary input files, and analyse the simulation results. 
We have considered three systems of different densities, containing 23, 107 and 207 W(CO)$_6$ molecules, which were randomly distributed within a simulation box with a side length of 20~nm.
The corresponding volume densities of the molecules were equal to $n \sim 2.9 \times 10^{-3}$~nm$^{-3}$, $\sim1.3 \times 10^{-2}$~nm$^{-3}$ and $\sim2.6 \times 10^{-2}$~nm$^{-3}$, respectively. Further details of the utilised simulation protocol are given in Section~\ref{sec:Simulation_protocol}.

\subsection{Interaction potentials}

The interatomic interactions in the simulated W(CO)$_6$ molecular systems and their fragments were described using the reactive CHARMM (rCHARMM) force field \cite{Sushko2016_rCHARMM} implemented in MBN Explorer. The rCHARMM framework enables simulations of systems with dynamically changing molecular topologies, which is essential for modelling chemical transformations in molecular and condensed matter systems, including those driven by irradiation. This methodology has been successfully employed in numerous previous studies of radiation-driven chemical processes underlying FEBID \cite{Sushko2016_IDMD, Prosvetov2021_BJN, DeVera2020, Prosvetov2022_PCCP, Prosvetov2023_EPJD, DeVera2019, Andreides_2023_JPCA_FeCO5, Lyshchuk_2025_JPCA_MeCpPtMe3}. 

To allow for the rupture and formation of covalent bonds, the radial part of the bonded interactions is described in rCHARMM by the Morse potential \cite{Sushko2016_rCHARMM}:
\begin{equation}
U_{\text{bond}}(r_{ij}) = D_{ij} \left[ e^{-2\beta_{ij}(r_{ij} - r_0)} - 2e^{-\beta_{ij}(r_{ij} - r_0)} \right] .
\label{eq:rCHARMM_bonded_pot}
\end{equation}
Here, $D_{ij}$ is the dissociation energy of the bond between atoms $i$ and $j$, $r_0$ is the equilibrium bond length, and $r_{ij}$ is the interatomic distance between atoms $i$ and $j$. The parameter $\beta_{ij} = \left( k_{ij}^{\text{bond}} / D_{ij} \right)^{1/2}$, where $k_{ij}^{\text{bond}}$ is the bond force constant, determines the steepness of the potential. 
Bonded interactions are truncated at a predefined cutoff distance, beyond which the bond is considered broken and the molecular topology of the system is updated accordingly. In the present simulations, this cutoff distance was set to 3.5~\AA~for all covalent bonds. This value is approximately equal to the sum of the van der Waals radii of atoms in the W(CO)$_6$ molecule, see Table~\ref{table:Lennard-Jones_parameters}.

The rupture of covalent bonds in the simulations automatically employs a modification of the potential functions for angular interactions:
\begin{equation}
U_{\text{angle}}(\theta_{ijk}) = 2 k^\theta _{ijk} \, \sigma(r_{ij}) \, \sigma(r_{jk}) \, [1 - \cos(\theta_{ijk} - \theta_0)] ,
\label{eq:rCHARMM_angular_pot}
\end{equation}
where $\theta _0$ is the equilibrium angle formed by atoms $i$, $j$ and $k$, and $k^\theta$ is the angle force constant. The sigmoid-type functions $\sigma(r_{ij})$ and $\sigma(r_{jk})$ account for the effect of bond breakage \cite{Sushko2016_rCHARMM}.

Non-bonded van der Waals interactions between atoms were described using the Lennard-Jones potential:
\begin{equation}
U(r_{ij}) = \varepsilon_{ij} \left[ 2\left( \frac{r_{\text{min}}}{r_{ij}} \right)^{6} - \left( \frac{r_{\text{min}}}{r_{ij}} \right)^{12} \right] , 
\label{eq:rCHARMM_LJ_potential}
\end{equation}
where $\varepsilon_{ij} = \sqrt{\varepsilon_i \varepsilon_j}$ and $r_{\rm min} = (r^i_{\rm min} + r^j_{\rm min})/2$.

\begin{table}[t]
\caption{
Parameters of the covalent bonded interaction, Eq.~(\ref{eq:rCHARMM_angular_pot}), for the W(CO)$_6$ parent molecule as well as CO, W(CO)$_n$ and WC(CO)$_n$ ($n = 1-5$) fragments. IDMD simulations using the rCHARMM force field allow changes in atom types and corresponding interaction parameters after the cleavage of a particular covalent bond and the formation of a specific fragment.}
\label{table:rCHARMM_param_parent}
\begin{tabular}{@{}llllll@{}}
\toprule
 Bond type & Atom types &  $r_0$ (\text{\AA})  &  $D_{ij}$ (kcal/mol) & $k_{ij}^{\text{bond}}$ (kcal/mol \text{\AA}$^{-1}$) & Species \\
\midrule
W--C(O) &  W--C    & 2.132   & 40.71   & 119.94 & parent W(CO)$_6$ molecule \\
 & WA--C$_5$    & 2.132   & 40.96   & 119.94 &  W(CO)$_5$, WC(CO)$_5$ \\
 & WA--C$_4$    & 2.132   & 42.17   & 119.94 &  W(CO)$_4$, WC(CO)$_4$ \\
 & WA--C$_3$    & 2.132   & 60.99   & 119.94 &  W(CO)$_3$, WC(CO)$_3$  \\
 & WA--C$_2$    & 2.132   & 49.15   & 119.94 &  W(CO)$_2$, WC(CO)$_2$  \\
 & WA--C$_1$    & 2.132   & 52.62   & 119.94 &  W(CO), WC(CO)  \\
W--C & WA--CA    & 2.132   & 124.91  & 119.94 &  WC(CO)$_5$ for the lone C \\
 & WA--CA    & 2.132   & 47.76  & 119.94 &  WC(CO)$_n$ ($n<5$) for the lone C \\
C--O & C--O    & 1.126   & 212.83  & 1493.91 & parent W(CO)$_6$ molecule  \\
  & C$_{1...5}$--O    & 1.126   & 212.83  & 1493.91 & all W-containing fragments  \\
  & CR--O    & 1.126   & 212.83  & 1493.91 & CO ligands  \\
\midrule
 & W--W    & 2.132   & 40.41   & 119.94 & W-containing molecular clusters  \\
 & O--O    & 1.126   & 212.83  & 1493.91 & O$_2$ molecule \\
\botrule
\end{tabular}
\end{table}

\begin{table}[t]
\caption{Parameters of the angular interaction, Eq.~(\ref{eq:rCHARMM_angular_pot}), for the W(CO)$_6$ parent molecule and its fragments.}
\begin{tabular}{@{}lllll@{}}
\toprule
 Angle &  $\theta _0$ (deg.)  &  $k^\theta$ (kcal/mol rad$^{-2}$) &  Species \\
\midrule
C--W--C    &  90.0   &  76.4   &  parent W(CO)$_6$ molecule  \\
W--C--O    &  180.0  &  14.0   &  parent W(CO)$_6$ molecule  \\
WA--C$_{1...5}$--O   &  180.0  &  14.0   &  all W-containing fragments  \\
\botrule
\end{tabular}
\label{table:rCHARMM_param_angular}
\end{table}

\begin{table}[t]
\caption{Parameters of the Lennard-Jones potential, Eq. (\ref{eq:rCHARMM_LJ_potential}), used to describe non-bonded interactions in the W(CO)$_6$ molecule and its fragments.}
\begin{tabular}{@{}lllll@{}}
\toprule
 Atom type &  $r_{\rm min}/2$ (\text{\AA})  &  $\varepsilon$ (kcal/mol) &  Ref. \\
\midrule
W, WA    &  1.76   &  0.1891   &  \cite{Oishi_2021_CompCondMatter, Wu_2024_JPCM}  \\
C, C$_{1...5}$, CA    &  1.95   &  0.0951  & \cite{Mayo1990}   \\
O    &  1.70   &  0.0957  &  \cite{Mayo1990}  \\
\botrule
\end{tabular}
\label{table:Lennard-Jones_parameters}
\end{table}

\begin{figure}[t]
\centering
\includegraphics[width=0.7\linewidth]{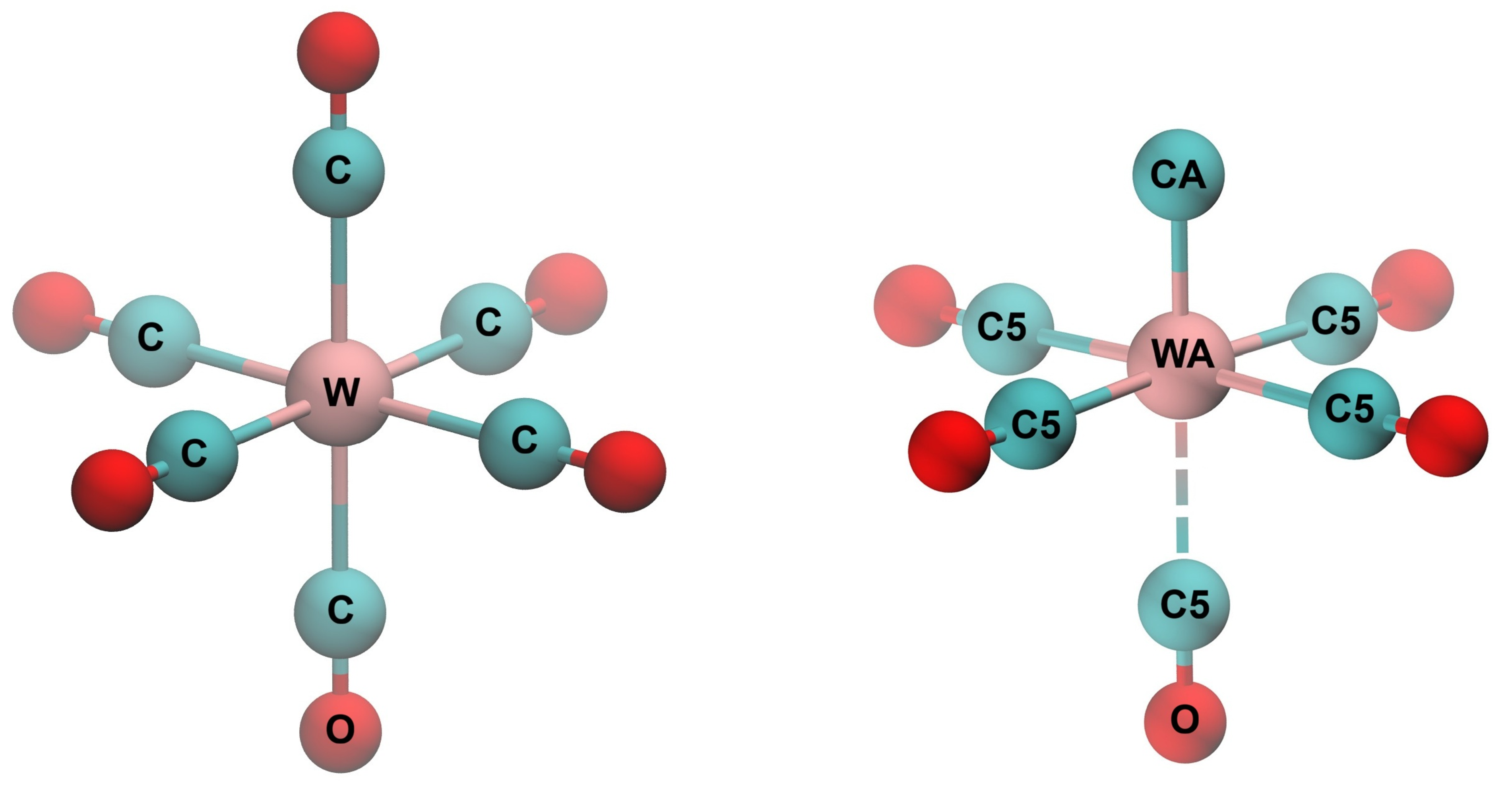} 
\caption{Change of atom types upon the cleavage of a C--O bond in the parent W(CO)$_6$ molecule (left panel) and the formation of a WC(CO)$_5$ fragment (right panel). The atom types for all the fragments considered in this study are summarised in  Table~\ref{table:rCHARMM_param_parent}. }
\label{fig:Figure_atom-types}
\end{figure}

The parameters of the bonded and angular interactions were previously determined from DFT calculations of potential energy scans for different covalent bonds in the parent W(CO)$_6$ molecule and its fragments W(CO)$_{6-n}$ and WC(CO)$_{5-n}$ \cite{DeVera2019}. 
In the cited study, the DFT calculations were performed using the Gaussian 09 software package and employed the hybrid B3LYP functional \cite{B3LYP} with the LanL2DZ  basis set \cite{lanl2dz_ref} for W atoms and the 6-31+G(d,p) basis set for C and O atoms. 
The corresponding parameter values for the different molecular species are listed in Table~\ref{table:rCHARMM_param_parent} and Table~\ref{table:rCHARMM_param_angular}. 
The corresponding atomic notations are discussed below and are also depicted in Figure~\ref{fig:Figure_atom-types} for the parent W(CO)$_6$ molecule and a WC(CO)$_5$ fragment.
The parameters for the non-bonded interactions for the parent W(CO)$_6$ molecule and its fragments are listed in Table~\ref{table:Lennard-Jones_parameters}.

Compared with the earlier MD study of W(CO)$_6$ fragmentation \cite{DeVera2019}, the present IDMD simulations employ a more detailed and comprehensive fragmentation model that allows changes in atom types and the corresponding interaction parameters based on information derived from the DFT calculations. 
In particular, we have accounted for changes in equilibrium bond lengths and dissociation energies of W--C and C--O bonds upon the formation of specific fragments (see Table~\ref{table:rCHARMM_param_parent}).
Further details of the utilised fragmentation model are provided in Appendix~\ref{secA1}.

The possibility of dynamically changing atom types for different molecular species within rCHARMM represents an advantage over other reactive force fields, including the widely used ReaxFF \cite{ReaxFF_Senftle_2016}. In ReaxFF, only a single atom type is typically assigned for each element, which requires a large number of parameters to describe bond-breaking and bond-formation processes.

In addition to fragmentation involving W--C bonds, the present study also explicitly enables C--O bond dissociation. This allows the simulation of fragmentation pathways leading to the formation of WC(CO)$_{5-n}$ species and other molecular products, such as O$_2$ molecules and larger molecular clusters.

\subsection{Irradiation conditions}
\label{sec:Methods_irradiation}

\begin{figure}
\centering
\includegraphics[width=0.7\linewidth]{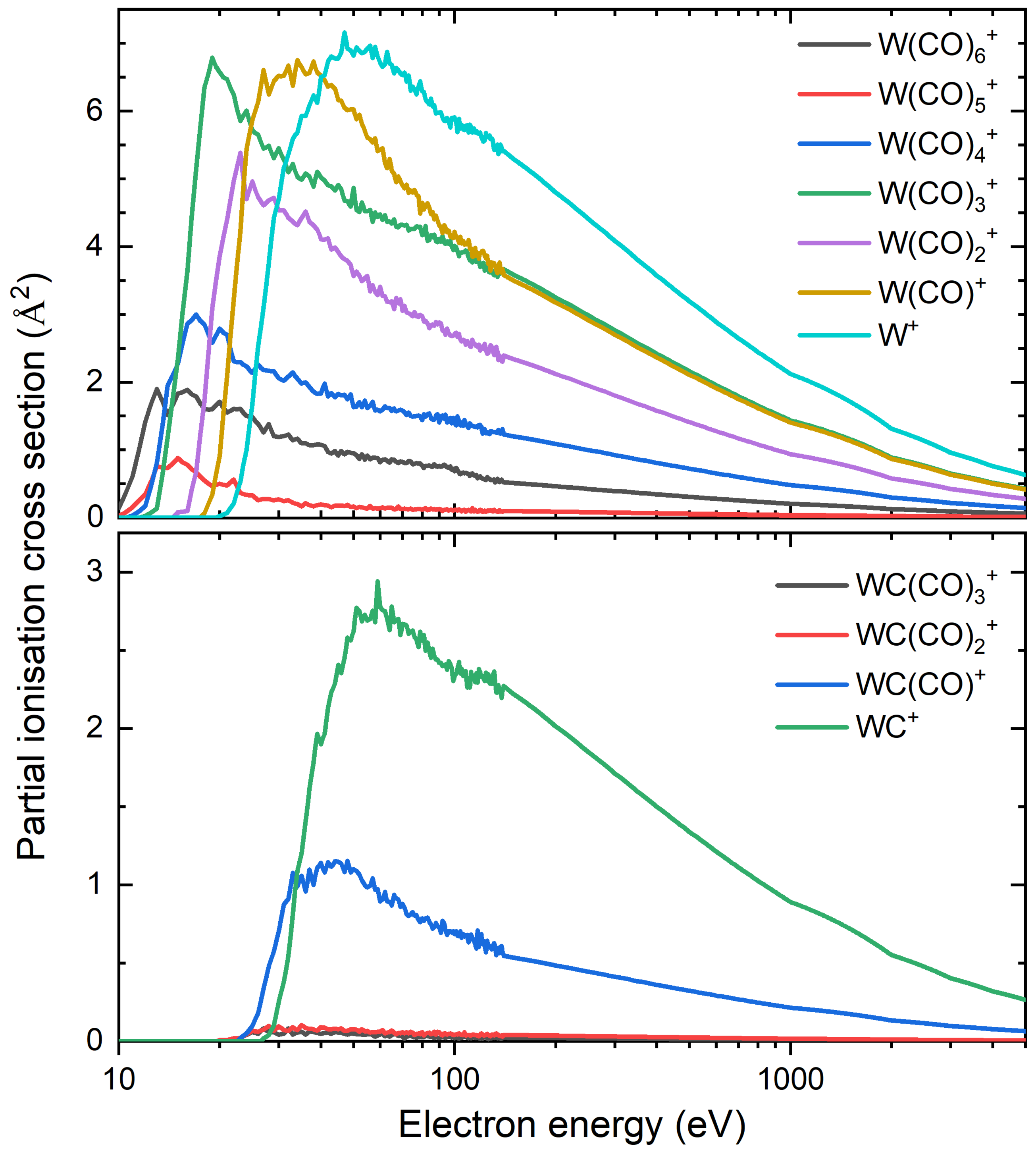} 
\caption{Absolute partial ionisation cross section (PICS) data for singly-charged cationic fragments W(CO)$_{6-n}^+$ ($n=0-6$) and WC(CO)$_{6-m}^+$ ($m=3-6$). The data have been obtained in the electron energy range from the ionisation threshold of the parent molecule ($\sim$8.5~eV) up to 5000~eV.}
\label{fig:Figure1_PICS_+}
\end{figure}

In this study, we consider dissociative ionisation (DI) as the primary mechanism governing the electron-induced fragmentation of W(CO)$_6$ molecules, for which experimental cross-section data are available \cite{WNOROWSKI-IJMS}. In the cited study, relative partial ionisation cross sections (PICS) for W(CO)$_6$ were reported for electron energies up to 140~eV. In the present study, absolute PICS values are obtained by combining these data with the total electron-impact ionisation cross section of W(CO)$_6$ calculated in Ref.~\cite{MEENUPANDEY}. In that work, the total ionisation cross section (TICS) was calculated using the ``complex scattering potential--ionisation contribution'' (CSP-ic) method \cite{CSP-IC}. First, the inelastic scattering cross section was calculated using the Spherical Complex Optical Potential (SCOP) method \cite{SCOPE}. Subsequently, the TICS for W(CO)$_6$ was derived from the inelastic channel by applying the CSP-ic formalism. By correlating these two datasets, we obtained absolute PICS data for different positively charged fragments of W(CO)$_6$ over a broad electron-energy range from the ionisation threshold of the parent molecule ($\sim$8.5~eV) up to 5000~eV, as shown in Figure~\ref{fig:Figure1_PICS_+}.

For electron energies up to 140 eV, the experimentally measured relative PICS data were normalised using the TICS according to the following normalisation condition:
\begin{equation}
\sigma^{\text{total}}_{\text{ion}}(E_i) = \sum_j \sigma^{\text{abs}}_j(E_i) \ .
\end{equation}
Here, $\sigma^{\text{total}}_{\text{ion}}(E_i)$ denotes the TICS at electron energy $E_i$, and $\sigma^{\text{abs}}_j(E_i)$ represents the absolute PICS for fragment $j$.
The absolute PICS values were obtained from the relative PICS data using the relation:
\begin{equation}
\sigma^{\text{abs}}_j(E_i) = 
\sigma^{\text{total}}_{\text{ion}}(E_i) \left[ \frac{\sigma^{\text{rel}}_j(E_i)}{\sum_k \sigma^{\text{rel}}_k(E_i)} \right] ,
\end{equation}
where $\sigma^{\text{rel}}_j(E_i)$ corresponds to the experimentally measured relative PICS. For electron energies above 140~eV, the relative PICS data were extrapolated and converted into absolute values using the same normalisation procedure. The calculated absolute PICS for W(CO)$_{6-n}^+$ ($n=0-6$) and WC(CO)$_{6-m}^+$ ($m=3-6$) species are presented in Figure~\ref{fig:Figure1_PICS_+}. 


In the performed IDMD simulations, a cubic simulation box with side length $d = 20$~nm and cross-sectional area $A = d^2$ was irradiated with a uniform electron beam of energy $E$. The electron flux density $J_0$ is given by:
\begin{equation}
J_0 = \frac{I_0}{e \times A} \ ,
\end{equation}
where $I_0$ is the electron beam current, and $e$ is the elementary charge.

For the low-density molecular systems considered in this study, the contribution of secondary electrons generated through ionisation events can be neglected in the first approximation. Under such conditions, the probability of multiple electron--molecule interactions within the simulation box is small, because the mean free path of electrons is one to two orders of magnitude larger than the characteristic system size. In addition, secondary electrons produced in ionisation events typically have low energies and therefore escape the system before inducing further reactions. As a result, the fragmentation dynamics are dominated by primary electron--molecule interactions.
Under this assumption, the fragmentation rate for a given covalent bond $i$ is given by:
\begin{equation}
P_i(r) = \sigma_i(E_j) \times J_0 \ ,
\end{equation}
where $\sigma_i(E_j)$ is the average fragmentation cross section for the $i^{\text{th}}$ bond (resulting in the formation of a specific fragment) induced by a primary electron with energy $E_j$.


In the present study, the fragmentation rates were evaluated at an incident electron energy of $E = 100$~eV, which lies within the range of electron energies ($\sim 70-100$~eV) typically used in gas-phase fragmentation experiments.   
%
Consequently, the absolute PICS values for the different fragmentation channels were taken at this energy to ensure consistency between the simulated electron--molecule interaction conditions and the cross-section data employed in the analysis. The ``default'' electron beam current and the simulation time were set to $I = 1$~$\mu$A and $t = 1$~ns, respectively, corresponding to a total electron fluence of $\Phi \approx 15.6$~nm$^{-2}$. 
The resulting electron fluence per W(CO)$_6$ molecule is comparable to that used in previous IDMD simulations of FEBID with W(CO)$_6$ \cite{Sushko2016_IDMD, DeVera2020}. However, in those studies, a significantly larger number of molecules were exposed to electron beam irradiation (typically, one to two orders of magnitude more than in the systems considered here), and the primary electron fluence per dwell time was higher by a corresponding factor.


The chosen value of fluence was kept constant throughout most of the simulations, unless stated otherwise (see Section \ref{sec:Results_varied-fluence}). To maintain this condition, the beam current and simulation time were adjusted accordingly for each simulation case. The results of this analysis are presented in Section~\ref{sec:Results_constant-fluence} and Section~\ref{sec:Post-fragmentation_chemistry}. In addition, the electron fluence was varied to investigate its effect on precursor fragmentation dynamics; these results are presented in Section~\ref{sec:Results_varied-fluence}.

\subsection{Simulation protocol}
\label{sec:Simulation_protocol}

The initial geometries of the systems with three different molecular densities (containing 23, 107, and 207 W(CO)$_6$ molecules) were generated using the built-in modeller tool of MBN Studio \cite{Sushko_2019_MBNStudio}.
In this procedure, a random distribution of molecules was generated based on the user-specified values of the volume number density of precursor molecules \cite{Sushko_2019_MBNStudio, MBNTutorials_50}. 
The resulting systems were first optimised using the velocity quenching algorithm. Subsequently, the systems were thermalised at 300~K for 1~ns using a Langevin thermostat with a damping time of 0.2~ps in order to generate equilibrated initial configurations for the subsequent IDMD simulations.



IDMD simulations of radiation-induced precursor fragmentation were performed for simulation times ranging from 1 to 30~ns for the system containing 23 molecules and from 1 to 10~ns for the denser systems. The electron beam current was adjusted accordingly in order to maintain a constant electron fluence across the simulations, as described in Section~\ref{sec:Methods_irradiation}. As a result, the beam current varied in the range from 1 $\mu$A to 33~nA. The highest beam current considered here lies within the range of beam currents used in previous IDMD simulations of FEBID \cite{Sushko2016_IDMD, DeVera2020, Prosvetov2021_BJN, Prosvetov2022_PCCP}. A summary of the simulations performed for the different parameter sets is given in Table~\ref{table: details of simulation}.

The IDMD simulations were performed with a time step of 0.5~fs. The molecular fragments produced at the end of the simulations were analysed, and the corresponding fragment appearance energies and their abundances were evaluated from this analysis.
Bond dissociation events were simulated by depositing a characteristic energy per electron-molecule collision, $E_{\rm dep}$, into the atoms forming the bond, thereby increasing their velocities \cite{Sushko2016_IDMD, DeVera2019}. In the present simulations, $E_{\rm dep}$ was set to 300~kcal/mol ($\sim$13~eV), which is consistent with the value used in previous IDMD simulations of FEBID with W(CO)$_6$ precursors \cite{DeVera2020}. In the cited study, this value of $E_{\rm dep}$ was selected by comparing the results of IDMD simulations of the electron irradiation-induced chemistry of W(CO)$_6$ precursors deposited on a surface (namely, the fraction of CO ligands remaining on a substrate covered with W(CO)$_6$ molecules as a function of electron dose) with the corresponding experimental data \cite{Rosenberg_2013_PCCP.15.4002}. 
For computational efficiency, all fragments produced during the simulations were assumed to be neutral, with the total electric charge of each fragment set to zero.



\begin{table}[h]
\caption{Summary of the IDMD simulations performed in this study.}
\label{table: details of simulation}
\begin{tabular}{ p{1.5cm} p{2.3cm} p{2.0cm} p{2.6cm} p{2.5cm} }
\hline
 Number of molecules &  Electron Fluence (nm$^{-2}$) & Simulation times (ns) & Beam currents & Number of independent runs  \\
\midrule
23     &   15.6   &  1--10   &  1 $\mu$A -- 100~nA  & 100 \\
23     &   15.6   &  15--30  &  66.7~nA -- 33.3~nA  & 20  \\
23     &   31.2   &  1--10   &  2 $\mu$A -- 200~nA  & 20 \\
23     &   62.4   &  1--10   &  4 $\mu$A -- 400~nA  & 20 \\
107    &   15.6   &  1--10   &  1 $\mu$A -- 100~nA  &  10 \\
207    &   15.6   &  1--10   &  1 $\mu$A -- 100~nA  & 10 \\
\botrule
\end{tabular}
\end{table}

\section{Results and Discussion}
\label{sec:Results}

In this section, we present the results of IDMD simulations of radiation-induced fragmentation and reactivity of gas-phase W(CO)$_6$ precursor molecules under different simulation conditions. The relative abundances of different fragments were determined from the number of species present at the end of each simulation divided by the initial number of parent molecules. To ensure statistically reliable results, multiple independent simulations were performed for each parameter set (see Table~\ref{table: details of simulation} above), and the reported quantities were obtained by averaging over these runs.

The relative fragment abundances provide insight into the dominant fragmentation pathways and the overall evolution of the irradiated molecular ensemble. In particular, analysing how these quantities depend on simulation time while maintaining a constant electron fluence allows us to assess whether the fragmentation dynamics exhibit a scaling behaviour with respect to irradiation parameters. If the fragment distributions remain invariant under different combinations of beam current and simulation time that correspond to the same total fluence, the system evolution can be considered fluence-controlled. In such a case, the irradiation-driven chemistry can be studied using shorter simulation times and correspondingly higher beam currents without affecting the final fragment distribution. On the other hand, deviations from this behaviour would indicate that the system dynamics depends explicitly on the irradiation parameters, requiring a more careful treatment of the simulation conditions.

\subsection{Fragmentation dynamics at different molecular densities}
\label{sec:Results_constant-fluence}

\begin{figure}
\centering
\includegraphics[width=1.0\linewidth]{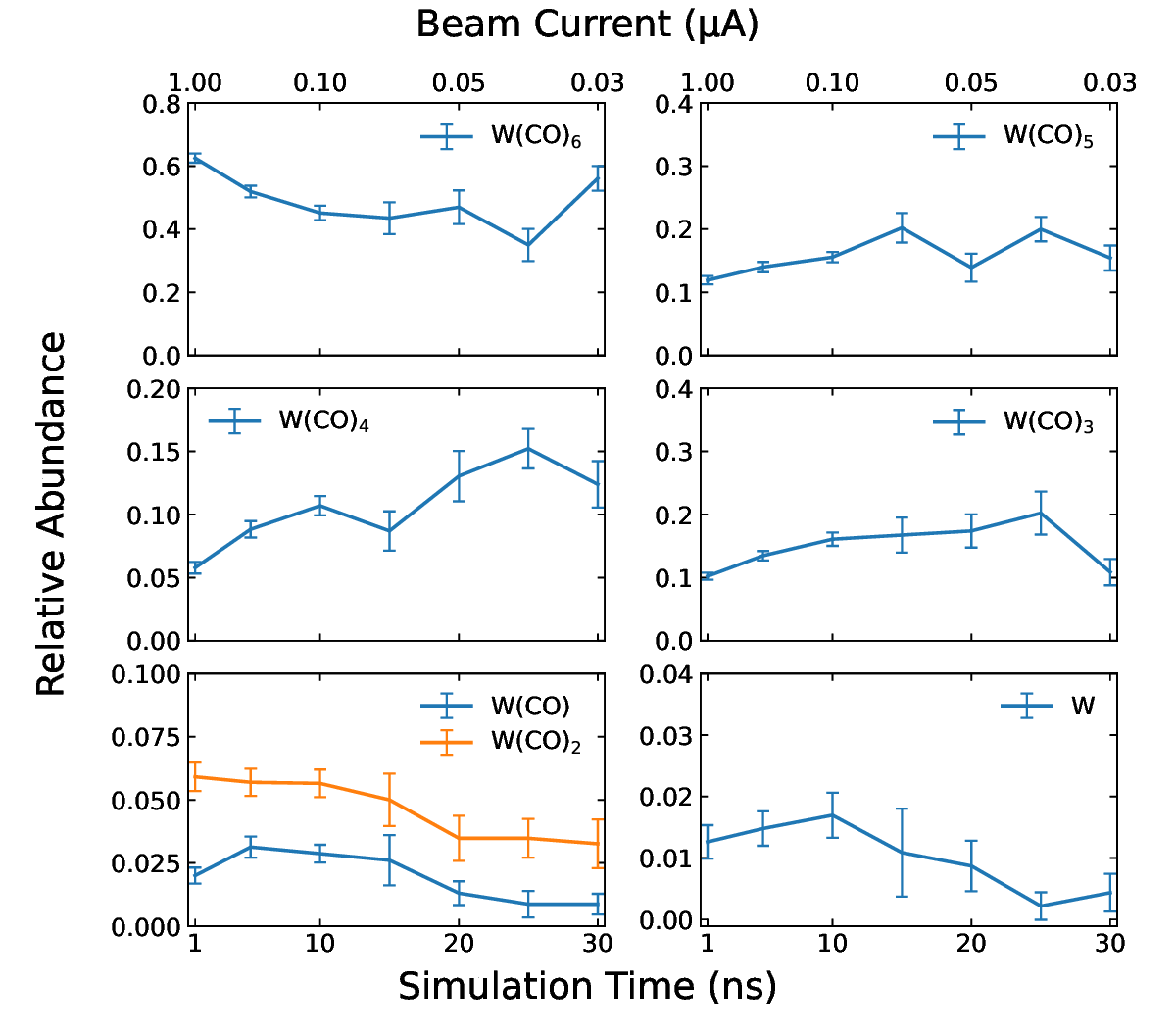}
\caption{The relative abundance of different W(CO)$_{6-n}$ ($n = 0-6$) molecular fragments at different simulation parameters for the low-density target comprising 23 parent molecules. The electron beam current (top x-axis) and simulation time (bottom x-axis) were rescaled to maintain a constant total electron fluence during each simulation. The error bars show the standard error for the respective number of independent simulations, as listed in Table~\ref{table: details of simulation}.}
\label{fig:Fragment_Abundance_23mol}
\end{figure}

We first analyse the system containing 23 W(CO)$_6$ molecules. Simulations were performed for irradiation times ranging from 1 to 30~ns in steps of 5~ns. For simulation times up to 10~ns, 100 independent runs were performed, whereas for longer simulation times 20 independent runs were carried out to maintain a reasonable computational cost while preserving statistical reliability. The results are summarised in Figure~\ref{fig:Fragment_Abundance_23mol}.

As shown in Figure~\ref{fig:Fragment_Abundance_23mol}, the relative abundance of the parent W(CO)$_6$ molecules initially decreases from $\sim$0.6 to $\sim$0.45 with increasing simulation time up to $\sim$10~ns, reflecting progressive fragmentation induced by electron irradiation. At longer simulation times and correspondingly smaller beam currents, the abundance of W(CO)$_6$ fluctuates around $\sim 0.4-0.5$. 
In contrast, the relative abundances of intermediate fragments W(CO)$_5$, W(CO)$_4$, and W(CO)$_3$ increase at simulation times up to $\sim$10~ns and subsequently fluctuate around $\sim 0.10-0.15$ at longer simulation times. These species, therefore, represent the dominant fragments in the irradiated ensemble under the present conditions.
The smaller fragments W(CO)$_2$, W(CO), and atomic W remain minor components of the molecular ensemble, with relative abundances on the order of one to several per cent. This behaviour indicates that, under the present low-density conditions, fragmentation channels involving the loss of more than three CO ligands are much less probable. 

The relatively small variation of the abundances for all W(CO)$_n$ species at simulation times above $\sim$10~ns suggests that the irradiated molecular ensemble has reached a state with a steady fragment distribution. For shorter simulation times and correspondingly higher beam currents, the fragmentation cascade has not yet reached such a regime, which is reflected in the higher abundance of small fragments [W(CO)$_2$, W(CO), and W] and the lower abundances of intermediate fragments [W(CO)$_5$, W(CO)$_4$, and W(CO)$_3$]. Therefore, for the considered electron fluence, IDMD simulations of the FEBID process performed with dwell times shorter than $\sim10$~ns would likely overestimate the production of highly fragmented species and, consequently, overestimate the tungsten content in the resulting deposits.

It should be noted that, although intra-ligand fragmentation processes involving C--O bond cleavage were allowed in the simulations, the formation of WC(CO)$_n$ species remains negligible for the conditions considered here. Cleavage of C--O bonds could in principle lead to the formation of free oxygen atoms, which may subsequently recombine with other O atoms or CO ligands to form species such as O$_2$ or CO$_2$. However, at the low molecular density, no formation of O$_2$ or CO$_2$ molecules was observed during the simulations. Such effects are discussed in Section~\ref{sec:Post-fragmentation_chemistry} for denser molecular ensembles.

\begin{figure}
\centering
\includegraphics[width=1.0\linewidth]{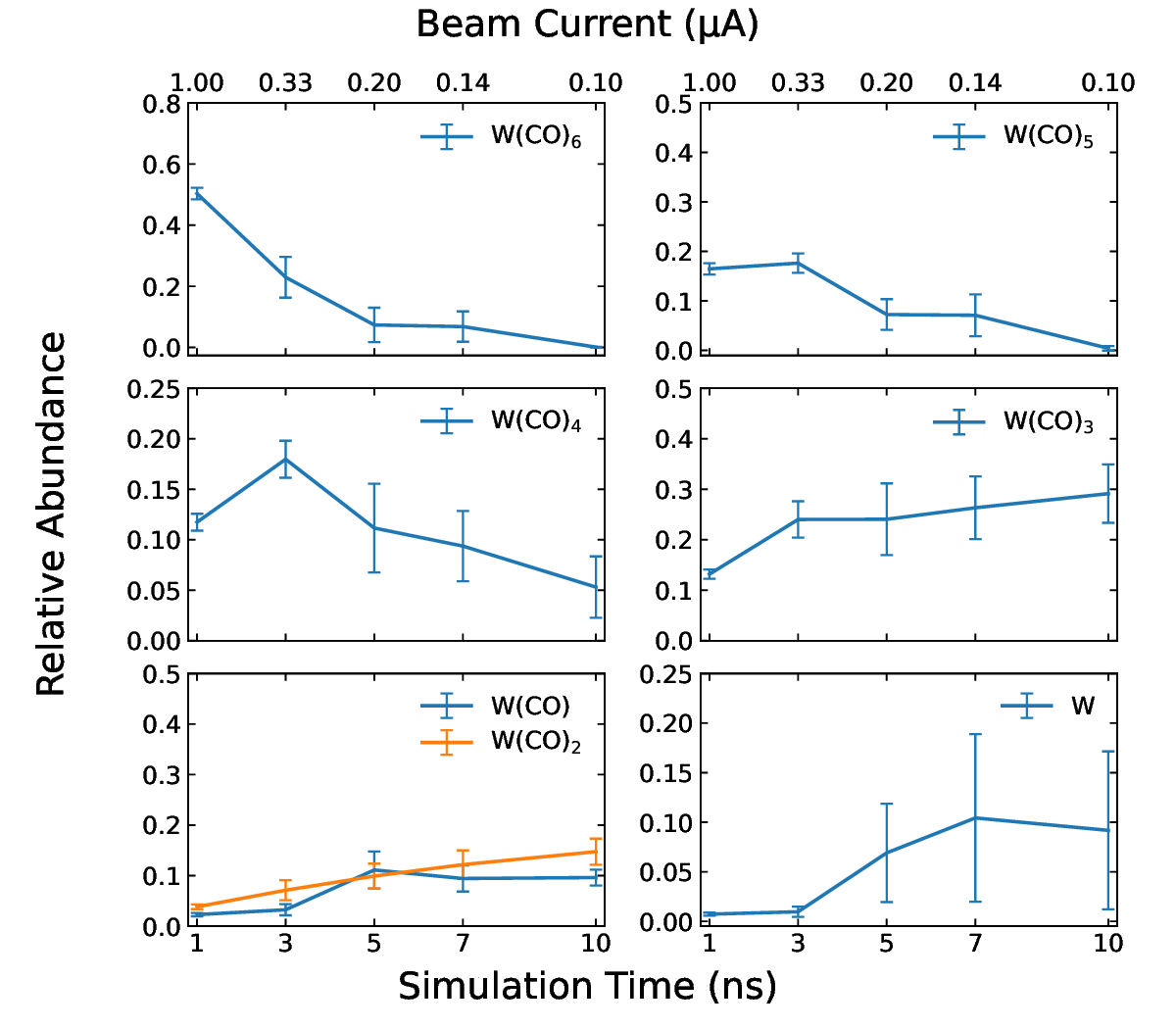}
\caption{The relative abundance of different W(CO)$_{6-n}$ ($n = 0-6$) molecular fragments at different simulation parameters for the case of the high-density target comprising 207 parent molecules. The electron beam current (top x-axis) and simulation time (bottom x-axis) were rescaled to maintain a constant total electron fluence during each simulation. 10 independent simulations were performed for each simulation time; the error bars show the corresponding standard error.}
\label{fig:Fragment_Abundance_207mol}
\end{figure}


The influence of molecular density on the fragmentation dynamics was studied by performing a similar set of simulations for a denser system containing 207 W(CO)$_6$ molecules. As in the low-density case described above, the total electron fluence was kept constant in all simulations. The results of this analysis are presented in Figure~\ref{fig:Fragment_Abundance_207mol}.   
Due to the increased computational cost associated with the larger system size, simulations were carried out for five irradiation times: 1, 3, 5, 7, and 10 ns. For each parameter set, the results were averaged over 10 independent simulation runs. 

\begin{figure}
\centering
\includegraphics[width=0.92\linewidth]{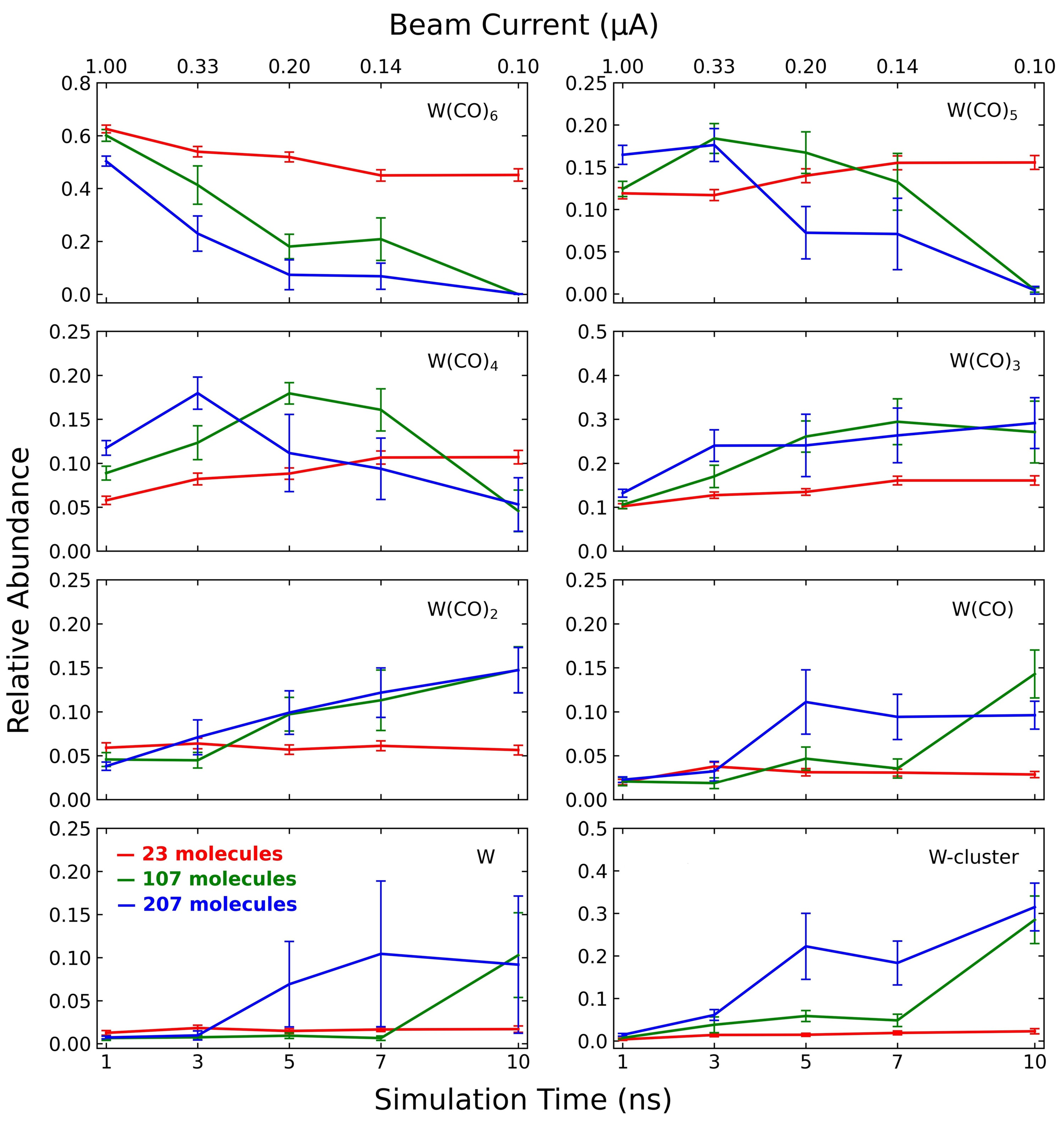}
\caption{The relative abundance of different W(CO)$_{6-n}$ ($n = 0-6$) molecular fragments at different simulation parameters for the case of three molecular densities, corresponding to 23 (red symbols), 107 (green symbols), and 207 (blue symbols) W(CO)$_6$ molecules placed in the simulation box with a side length of 20~nm. The electron fluence has been kept constant throughout all simulations.}
\label{fig:density effect}
\end{figure}

The fragmentation behaviour at higher molecular density differs from that observed for the lower-density system. The relative abundances of W(CO)$_6$ and W(CO)$_5$ species decrease steadily with increasing simulation time and approach zero at a simulation time of 10~ns. A similar behaviour is observed for W(CO)$_4$, although a small residual fraction of these species (about 5\% of the initial number of W(CO)$_6$ molecules) remains at the longest simulation time.

In contrast, the abundances of the W(CO)$_3$, W(CO)$_2$, W(CO), and W fragments increase gradually with increasing simulation time and eventually approach approximately constant values at simulation times above $\sim$5~ns. The steady-state abundances differ for each fragment, with W(CO)$_3$ being the most abundant species in the ensemble, while the other small fragments exhibit abundances approximately threefold lower.


To further investigate the influence of the molecular environment on fragmentation dynamics, an additional set of simulations was performed for an intermediate-density system comprising 107 W(CO)$_6$ molecules. All irradiation parameters were kept identical to the simulations described above. 
Figure~\ref{fig:density effect} compares the relative abundances of fragments for different simulation times and corresponding electron beam currents for the three systems containing 23, 107, and 207 W(CO)$_6$ molecules.
%
The figure clearly demonstrates that the molecular density of the system influences the fragmentation dynamics. The temporal evolution patterns and the final fragment distributions for the systems containing 107 and 207 molecules are similar and differ noticeably from those observed for the low-density system. In particular, the relative abundances of the parent W(CO)$_6$ molecule and the W(CO)$_5$ and W(CO)$_4$ fragments decrease more rapidly with increasing molecular density.

In contrast, the relative abundances of the smaller fragments exhibit a slight increase with increasing molecular density. This trend indicates that higher molecular densities promote more extensive fragmentation of the precursor molecules, resulting in more frequent formation of intermediate and smaller species, particularly W(CO)$_3$. This behaviour can be attributed to an increased probability of intermolecular collisions and interactions involving fragments, free CO ligands and neighbouring molecules at higher densities, which can facilitate additional dissociation pathways. 
These results demonstrate that, in contrast to the low-density case, the fragmentation dynamics at higher densities is influenced not only by direct electron-induced processes, but also by secondary chemical reactions involving fragments.

\subsection{Variation of electron fluence}
\label{sec:Results_varied-fluence}

\begin{figure}
\centering
\includegraphics[width=0.95\linewidth]{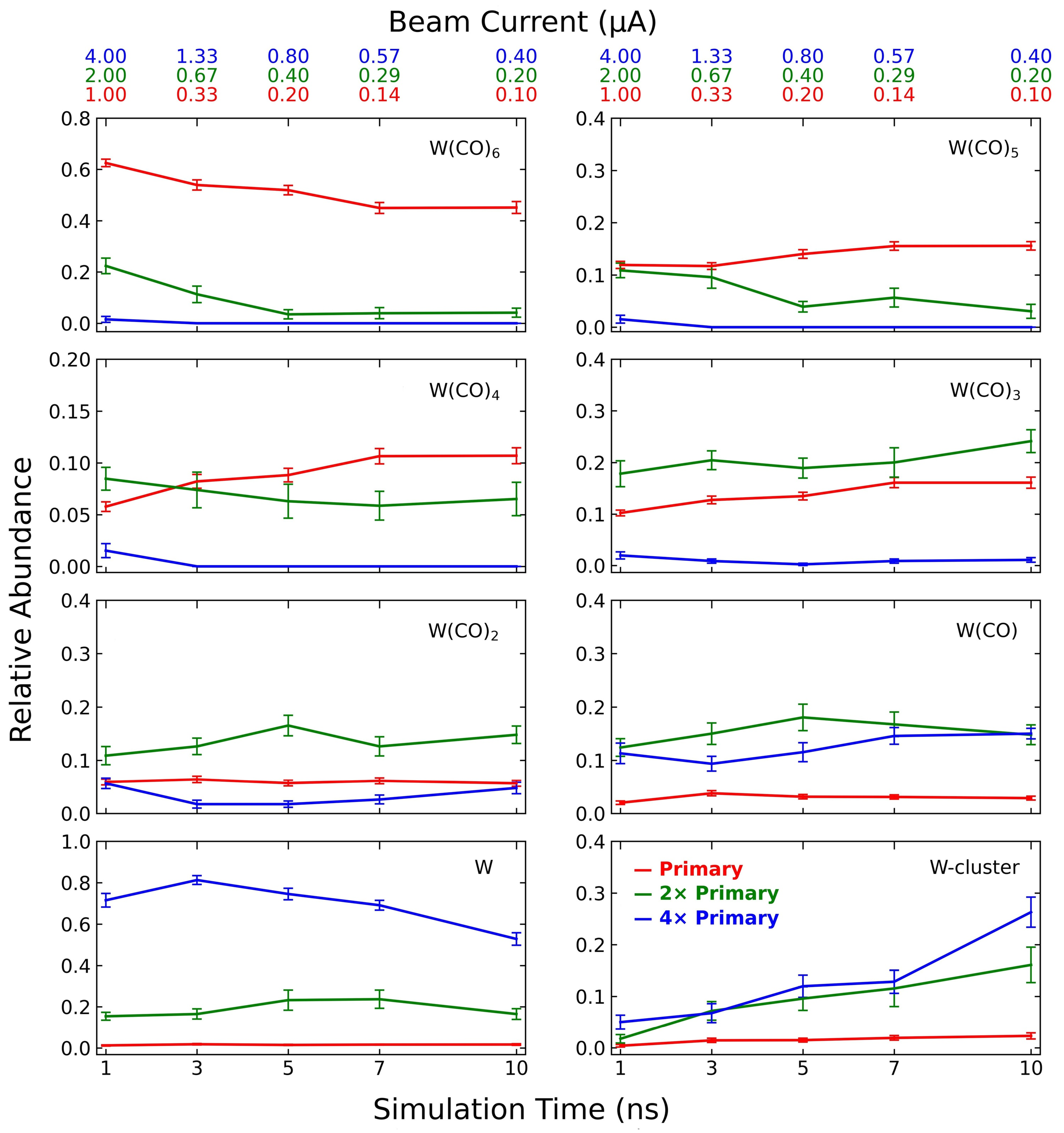}
\caption{The relative abundances of W(CO)$_6$ fragments at a reference electron fluence of $\Phi \approx 15.6$~nm$^{-2}$ (red symbols) and at two- and four-fold higher fluences (green and blue symbols, respectively) for a system containing 23 W(CO)$_6$ molecules. 20 independent simulations were performed for each simulation time; the error bars show the corresponding standard error.}
\label{fig:fluency effect}
\end{figure}

In the simulations described in the previous section, the electron fluence was kept constant throughout all simulations. To investigate the influence of this parameter on the precursor fragmentation dynamics, additional simulations with systematically varied fluence were performed using the low-density system containing 23 W(CO)$_6$ molecules as an illustrative case. Simulations were carried out at electron fluences equal to the reference value used in the simulations discussed above ($\Phi \approx 15.6$ electrons per nm$^{2}$), as well as at two- and fourfold higher values. For each case, the total electron fluence was kept constant for different combinations of simulation time and the corresponding beam current. For each set of parameters, 20 independent simulations were performed. The results of this analysis are presented in Figure~\ref{fig:fluency effect}.

Figure~\ref{fig:fluency effect} demonstrates that the electron fluence has a pronounced effect on the fragmentation dynamics of the precursor molecules. As the electron fluence increases, the fragmentation rate increases accordingly. This behaviour is expected, since a higher fluence corresponds to a larger number of electron--molecule interaction events, thereby increasing the probability of electron-induced bond dissociation. Consequently, increasing the electron fluence shifts the fragment distribution towards smaller molecular species. This trend is particularly evident at the highest fluence considered (blue symbols in Figure~\ref{fig:fluency effect}), where the system rapidly evolves towards a state dominated by W(CO) fragments and atomic W.

The increased abundance of highly fragmented tungsten-containing species at higher electron fluence also has significant implications for the subsequent chemical evolution of the system. In particular, the formation of under-coordinated fragments increases the probability of intermolecular interactions, which can lead to the aggregation of tungsten atoms and small W(CO)$_n$ fragments and, ultimately, to the formation of W-containing molecular clusters. 
%
The following section analyses in greater detail the formation and growth of W-containing molecular clusters under different irradiation conditions.

\subsection{Post-fragmentation chemical transformations}
\label{sec:Post-fragmentation_chemistry}


An important consequence of irradiation-induced fragmentation of W(CO)$_6$ molecules is the subsequent aggregation of tungsten-containing fragments, which leads to the formation of W-rich nanoclusters. The ability to control the formation of such clusters through irradiation parameters, such as electron fluence and beam current, is of particular interest, because it directly relates to the mechanisms governing nanostructure formation under electron irradiation.

To analyse this process, we evaluated the relative abundance of W atoms that are present in clusters of different sizes (i.e. in all species containing a larger number of atoms than a parent W(CO)$_6$ molecule) as a function of simulation time for the precursor densities and electron fluences discussed in the previous sections. 

The bottom-right panel of Figure~\ref{fig:fluency effect} shows the relative abundance of W atoms in formed clusters for different electron fluences discussed in Section~\ref{sec:Results_varied-fluence}.
These results indicate that increasing the electron fluence significantly enhances the formation of W-containing aggregates. Higher fluences lead to more extensive fragmentation of precursor molecules and consequently generate a larger number of reactive tungsten-containing fragments capable of interacting with each other and forming molecular clusters. In the simulations performed for 10~ns, the resulting cluster structures incorporate up to approximately 30\% of the tungsten atoms that were initially present in the system in the form of W(CO)$_6$ molecules (see blue symbols in Figure~\ref{fig:fluency effect}).

The bottom-right panel of Figure~\ref{fig:density effect} shows the relative abundance of W atoms in such objects for the three precursor densities discussed in Section~\ref{sec:Results_constant-fluence}.
The results demonstrate a clear density dependence of cluster formation. At the lowest molecular density considered (red symbols in Figure~\ref{fig:density effect}), the abundance of W atoms in such clusters remains negligible (around 2\%). As the precursor density increases, a larger fraction of tungsten atoms becomes incorporated into molecular clusters during the simulation. This behaviour can be attributed to the increased probability of collisions between fragmentation products at higher molecular densities, which facilitates cluster formation.

\begin{figure}[t!]
\centering
\includegraphics[width=0.8\linewidth]{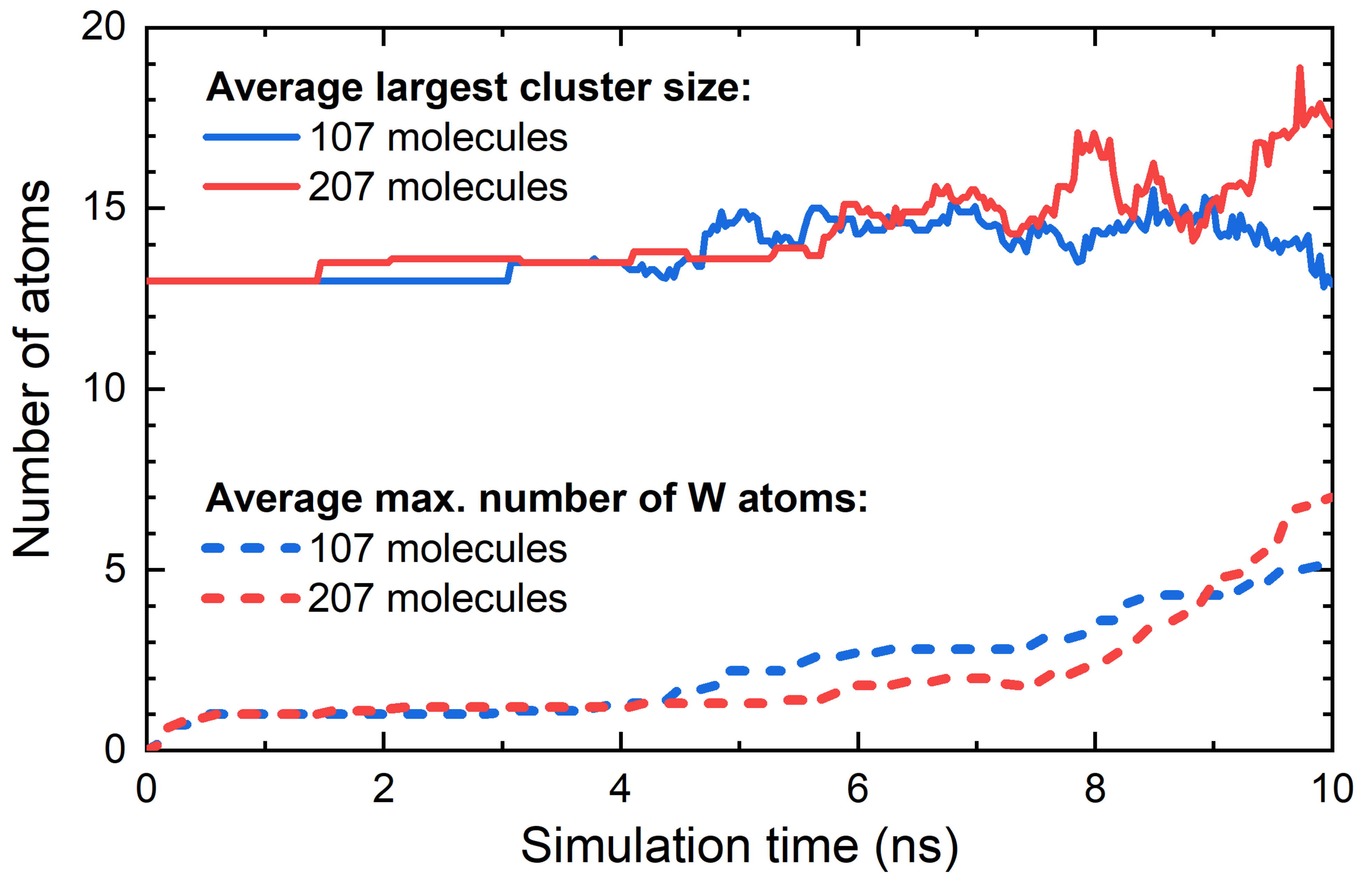}
\caption{Time evolution of the average largest cluster size (solid lines) and the average maximum number of tungsten atoms in a cluster (dashed lines) for the systems initially containing 107 and 207 W(CO)$_6$ molecules. The curves represent the averaged values obtained from 10 independent simulation runs.}
\label{fig:cluster dynamics}
\end{figure}

We also analysed the structural evolution of the clusters formed during the simulations. For this purpose, we quantified the evolution of the maximal cluster size (defined as the number of atoms in the largest species present in the simulation box) and the largest number of tungsten atoms of type WA (which is present in all fragments but not in the parent molecule, see Table~\ref{table:rCHARMM_param_parent}) contained within a single species during the 10~ns simulations for two illustrative systems containing 107 and 207 molecules. The results were averaged over 10 independent simulation runs. The corresponding data are presented in Figure~\ref{fig:cluster dynamics}.

At the beginning of the simulations, the number of tungsten atoms of type WA is equal to zero (see the dashed curves in Figure~\ref{fig:cluster dynamics}), indicating that no fragments are present in the simulation box and that the system initially contains only intact W(CO)$_6$ molecules. As the simulation continues, the number of WA atoms increases and remains equal to one up to $\sim$4~ns, reflecting the formation of individual W(CO)$_n$ fragments. At longer simulation times, this quantity begins to increase, indicating the merging of several fragments into molecular clusters containing more than one tungsten atom. At the end of 10~ns simulations, the formed clusters contain, on average, between 5 and 7 tungsten atoms, depending on the molecular density of the target.

The solid curves in Fig.~\ref{fig:cluster dynamics} show the evolution of the number of atoms (of all types) in the largest structure present in the system at each instant. As expected, the initial value of this quantity is 13, corresponding to intact W(CO)$_6$ molecules. During the simulations, this number gradually increases and exhibits fluctuations due to radiation- and collision-induced bond breakage and bond formation events occurring in the system. Clusters are initially formed by the merging of individual W(CO)$_n$ fragments, after which these structures may further combine with other fragments or clusters to form larger W-containing aggregates.

\begin{figure}[t!]
\centering
\includegraphics[width=0.72\linewidth]{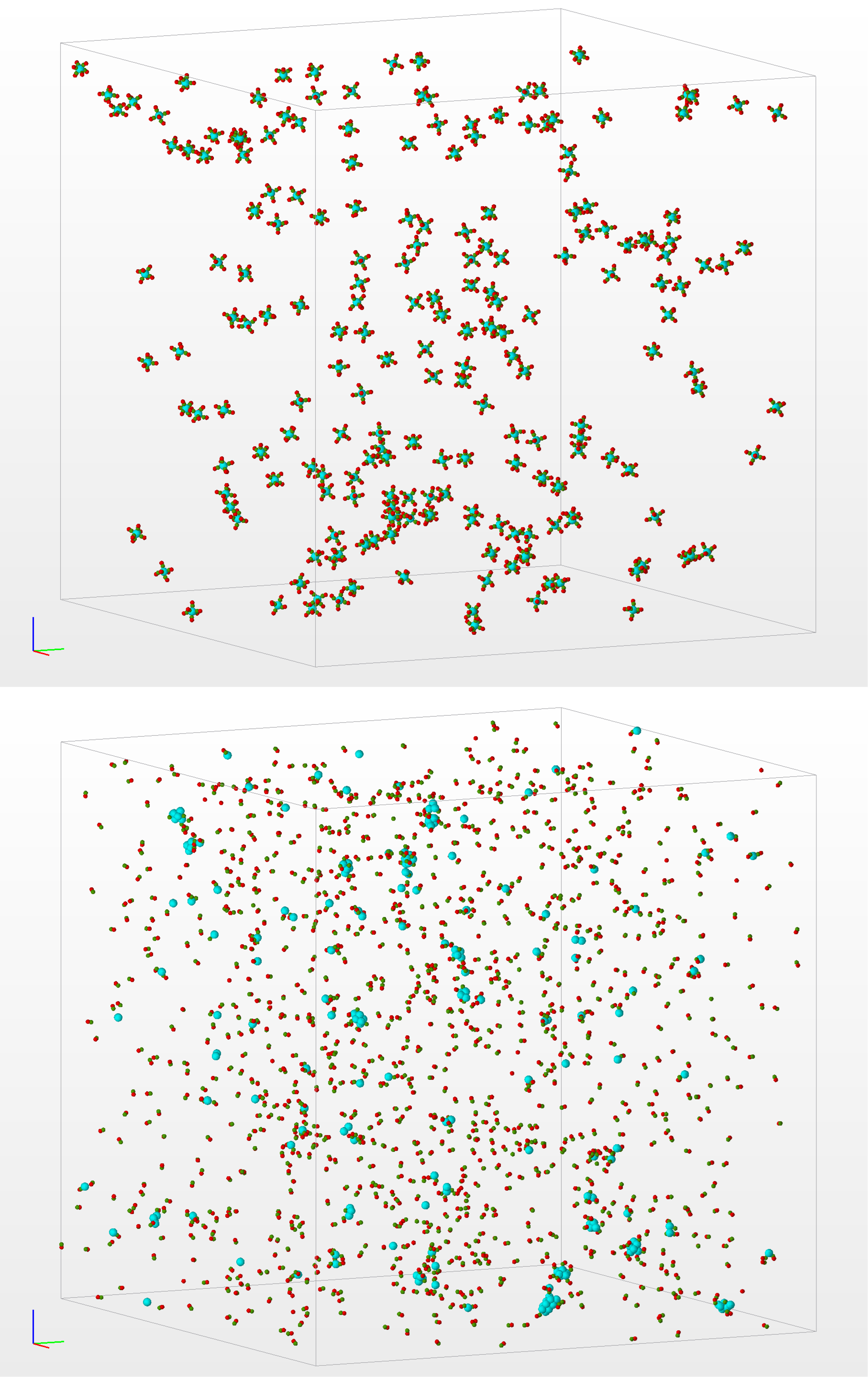}
\caption{Visualisation of the radiation-induced fragmentation of W(CO)$_6$ molecules and the formation of small W-rich molecular clusters. The top panel shows the initial configuration of the system, in which 207 W(CO)$_6$ molecules are uniformly distributed within the simulation box. The bottom panel shows the final state of the system after the 10~ns simulation. Tungsten atoms are depicted as light-blue spheres, while carbon and oxygen atoms are shown as green and red spheres, respectively.}
\label{fig:visualization}
\end{figure}

In the case of the denser system containing 207 molecules (see the solid red curve in Figure~\ref{fig:cluster dynamics}), the largest structures contain, on average, 17-18 atoms. However, the simulated system exhibits a distribution of W-containing clusters with different sizes and compositions. Representative clusters observed in the simulations contain up to 29 atoms, while the tungsten content in some structures can reach $\sim65-70$~at.\%, for example, in clusters such as W$_{12}$C$_3$O$_3$ and W$_8$C$_2$O (see an illustrative simulation snapshot in the bottom panel of Figure~\ref{fig:visualization}).
It should be emphasised that C–O bond cleavage was explicitly included in the present simulations; therefore, the numbers of carbon and oxygen atoms in the formed clusters are not necessarily equal, which leads to structures with a reduced oxygen content. 


Figure~\ref{fig:visualization} shows two simulation snapshots illustrating the cluster formation process. The top panel corresponds to the initial configuration of the system, in which 207 W(CO)$_6$ molecules are uniformly distributed within the simulation box prior to irradiation. The bottom panel illustrates the final state of the system after the 10~ns simulation. In this configuration, the precursor molecules have undergone extensive fragmentation, and the resulting species have recombined and aggregated to form clusters of different sizes. The formed structures represent mixed organometallic molecular clusters in which several tungsten atoms (shown as light-blue spheres) are coordinated by surrounding carbon and oxygen atoms (shown as green and red spheres, respectively).


\begin{figure}
\centering
\includegraphics[width=0.75\linewidth]{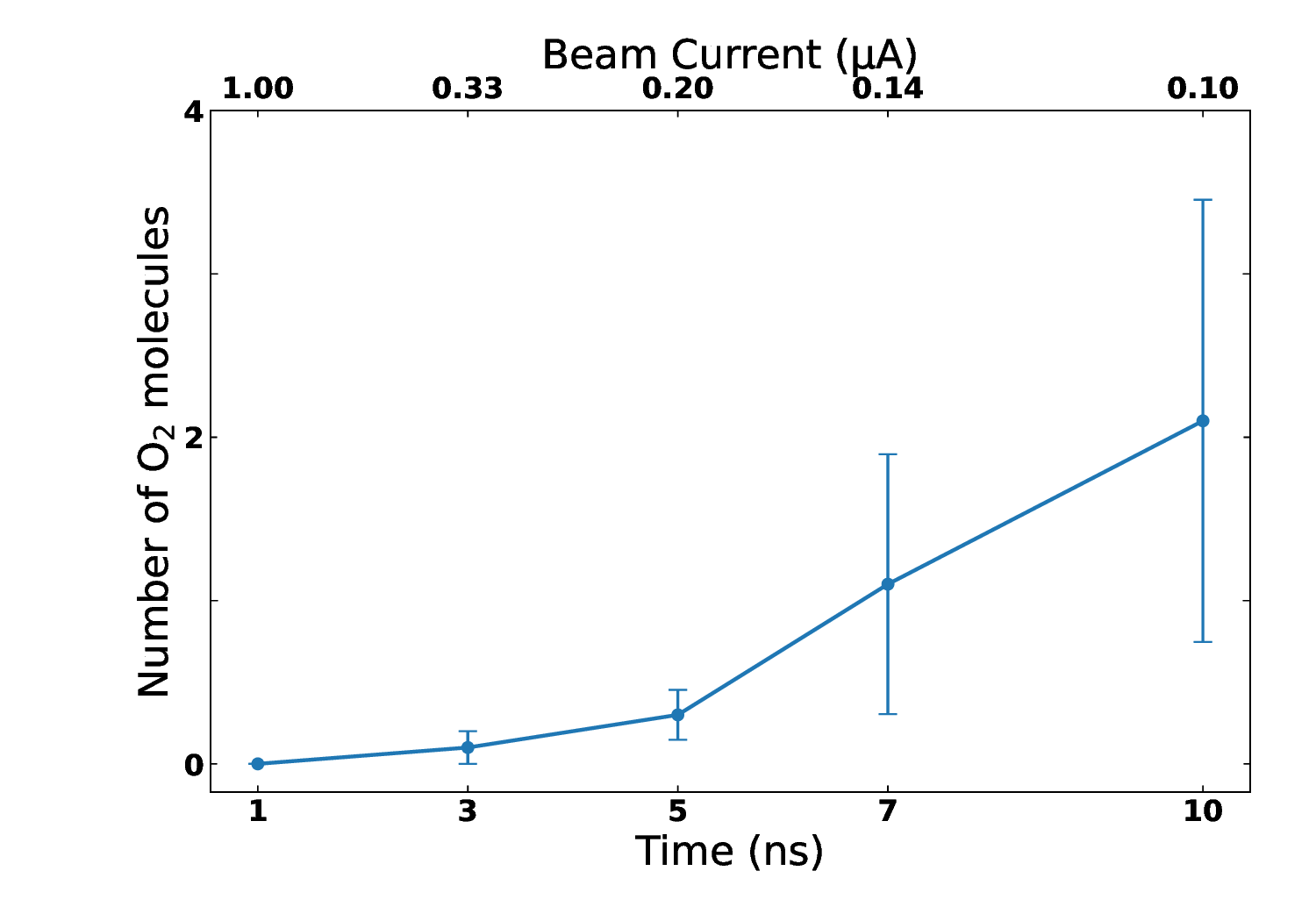}
\caption{Average number of O$_2$ formed during the IDMD simulations at indicated simulation times and beam currents. The studied system initially contained 207 W(CO)$_6$ molecules. 10 independent simulations were performed for each simulation time; the error bars show the corresponding standard error.}
\label{fig:O2_abundance}
\end{figure}

In addition to tungsten-containing clusters, the formation of O$_2$ molecules was also observed in simulations performed at the highest molecular density. Figure~\ref{fig:O2_abundance} shows the average number of O$_2$ molecules formed as a function of simulation time. The formation of O$_2$ originates from recombination reactions involving oxygen atoms released during C--O bond cleavage in the precursor molecules. However, the overall abundance of O$_2$ remains low under the investigated conditions. Similar behaviour was observed in simulations performed at higher electron fluence and longer irradiation times, where the number of O$_2$ molecules formed remains negligible compared with the dominant tungsten-containing fragments and clusters.

\section{Conclusion}
\label{sec:Conclusions}

In this work, the electron-induced fragmentation dynamics of W(CO)$_6$ precursor molecules has been investigated computationally using irradiation-driven molecular dynamics (IDMD) simulations. The simulations incorporated the experimentally derived partial ionisation cross sections for W(CO)$_n$ and WC(CO)$_n$ fragments together with the calculated total ionisation cross section for the parent W(CO)$_6$ molecule, enabling an accurate quantitative description of bond breaking and subsequent chemical reactions in the molecular system exposed to electron irradiation.

The obtained results demonstrate that electron irradiation leads to progressive fragmentation of W(CO)$_6$ molecules into a series of W(CO)$_n$ fragments accompanied by the release of CO ligands. The relative abundances of the different fragments evolve systematically with irradiation time and electron fluence. In particular, the population of the parent W(CO)$_6$ molecule decreases steadily, while smaller fragments such as W(CO)$_3$, W(CO)$_2$, W(CO), and atomic W become increasingly dominant at longer irradiation times and higher electron fluences.

The simulations further reveal that precursor density and electron fluence play an important role in determining the overall fragmentation dynamics. Higher precursor densities increase the probability of interactions between fragmentation products, thereby promoting aggregation processes and leading to enhanced formation of tungsten-rich clusters. Similarly, increasing electron fluence accelerates fragmentation and results in a larger fraction of tungsten atoms incorporated into clusters.

The formation of W-containing molecular clusters is identified as one of the key outcomes of the irradiation process. Cluster growth proceeds through the aggregation of tungsten-containing fragments produced during precursor decomposition. Analysis of the cluster size evolution indicates that the aggregation dynamics depends on both the irradiation conditions and the molecular density of the system.

Overall, the present study provides detailed molecular-level insight into the mechanisms governing electron-induced fragmentation and subsequent aggregation processes in W(CO)$_6$ precursor systems. The obtained results contribute to a better understanding of the fundamental processes underlying electron-beam-induced nanostructure formation and may guide the selection of simulation parameters for future IDMD simulations of the FEBID process.

\bmhead{Acknowledgments}

This work was supported by the COST Action CA20129 (MultIChem) and the COST Innovators Grant IG20129 (INDICO), both funded by the European Cooperation in Science and Technology. The possibility of performing computer simulations at the Goethe-HLR cluster of the Frankfurt Center for Scientific Computing is gratefully
acknowledged.

\section*{Author contribution statement}
Soumyo Kheto: Data curation, formal analysis, investigation, visualisation, writing – original draft.
Alexey Verkhovtsev: Conceptualisation, data curation, investigation, methodology, visualisation, writing – original draft, writing – review {\&} editing. 
Bobby Antony: Conceptualisation, Methodology, Resources, Supervision, Writing – review {\&} editing.
Andrey V. Solov'yov: Conceptualisation, investigation, methodology, project administration, resources, supervision, writing – review {\&} editing.

\section*{Data availability}

The data supporting this article have been included as part of the Supplementary Information.
\begin{appendices}

\section{W(CO)$_6$ fragmentation model}
\label{secA1}

Here, we provide a brief description of the main principles of the molecular fragmentation model used in the performed IDMD simulations. Table~\ref{table:Appendix_reactions} provides information on the chemical reactions and specific reaction channels included in the simulation protocol.


The loss of CO ligands is assumed to occur sequentially through the formation of a W(CO)$_{n-1}$ product from a W(CO)$_n$ species. Simultaneous loss of multiple CO ligands, corresponding to the formation of W(CO)$_{n-m}$ ($m>1$) products from a W(CO)$_n$, is energetically unfavourable and, therefore, is not considered in the utilised fragmentation model.

If a C--O bond is broken, leading to the formation of a WC(CO)$_n$ product, further fragmentation proceeds with a high probability through the loss of CO ligands, as the W--C(O) bonds are much weaker than the C--O bonds and the W--C bond without an oxygen atom (see Table~\ref{table:rCHARMM_param_parent}).

The formation of a W(CO)$_n$ species is assumed to occur through the recombination of a W(CO)$_{n-1}$ ligand with a free CO radical. The simultaneous recombination of several $m$CO radicals ($m>1$) with a W(CO)$_{n-m}$ fragment has not been considered due to the low probability of several CO ligands being located in close proximity at any given time. 


\begin{table}[t]
\caption{Chemical reactions and specific reaction channels included in the IDMD simulation protocol. W and C denote tungsten and carbon atoms in the parent W(CO)$_6$ molecule. WA and C$_{1...5}$ denote those in W(CO)$_n$ and WC(CO)$_n$ fragments. CR denotes a carbon atom in the CO ligand.}
\label{table:Appendix_reactions}
\begin{tabular}{p{4.2cm}|p{3.3cm}|p{4.1cm}}
\hline
\textbf{Reaction} & \textbf{Initial atom types} & \textbf{Final atom types} \\
\midrule

W(CO)$_6$ $\rightarrow$ W(CO)$_5$ + CO
& W, 6$\times$C, 6$\times$O
& WA, 5$\times$C5, 5$\times$O; CR, O \\

W(CO)$_5$ $\rightarrow$ W(CO)$_4$ + CO
& WA, 5$\times$C5, 5$\times$O
& WA, 4$\times$C4, 4$\times$O; CR, O \\

W(CO)$_4$ $\rightarrow$ W(CO)$_3$ + CO
& WA, 4$\times$C4, 4$\times$O
& WA, 3$\times$C3, 3$\times$O; CR, O \\

W(CO)$_3$ $\rightarrow$ W(CO)$_2$ + CO
& WA, 3$\times$C3, 3$\times$O
& WA, 2$\times$C2, 2$\times$O; CR, O \\

W(CO)$_2$ $\rightarrow$ W(CO) + CO
& WA, 2$\times$C2, 2$\times$O
& WA, C1, O; CR, O \\

\midrule

W(CO)$_6$ $\rightarrow$ WC(CO)$_5$ + O
& W, 5$\times$C, CA, 6$\times$O
& WA, 5$\times$C5, CA, 5$\times$O; CR, O \\

WC(CO)$_5$ $\rightarrow$ WC(CO)$_4$ + CO
& WA, 5$\times$C5, CA, 5$\times$O
& WA, 4$\times$C4, CA, 4$\times$O; CR, O \\

WC(CO)$_4$ $\rightarrow$ WC(CO)$_3$ + CO
& WA, 4$\times$C4, CA, 4$\times$O
& WA, 3$\times$C3, CA, 3$\times$O; CR, O \\

WC(CO)$_3$ $\rightarrow$ WC(CO)$_2$ + CO
& WA, 3$\times$C3, CA, 3$\times$O
& WA, 2$\times$C2, CA, 2$\times$O; CR, O \\

WC(CO)$_2$ $\rightarrow$ WC(CO) + CO
& WA, 2$\times$C2, CA, 2$\times$O
& WA, C1, CA, O; CR, O \\

WC(CO) $\rightarrow$ WC + CO
& WA, C1, CA, O
& WA, CA; CR, O \\

\midrule

W(CO)$_5$ + CO $\rightarrow$ W(CO)$_6$
& WA, 5$\times$C5, 5$\times$O; CR, O
& W, 6$\times$C, 6$\times$O \\

W(CO)$_4$ + CO $\rightarrow$ W(CO)$_5$
& WA, 4$\times$C4, 4$\times$O; CR, O
& WA, 5$\times$C5, 5$\times$O \\

W(CO)$_3$ + CO $\rightarrow$ W(CO)$_4$
& WA, 3$\times$C3, 3$\times$O; CR, O
& WA, 4$\times$C4, 4$\times$O \\

W(CO)$_2$ + CO $\rightarrow$ W(CO)$_3$
& WA, 2$\times$C2, 2$\times$O; CR, O
& WA, 3$\times$C3, 3$\times$O \\

W(CO) + CO $\rightarrow$ W(CO)$_2$
& WA, C1, O; CR, O
& WA, 2$\times$C2, 2$\times$O \\

W + CO $\rightarrow$ W(CO)
& WA; CR, O
& WA, C1, O \\
\botrule
\end{tabular}
\end{table}

\end{appendices}



\bibliography{sn-bibliography}

@article{Utke2008,
author = {Utke, I. and Hoffmann, P. and Melngailis, J.},
title = {{Gas-Assisted Focused Electron Beam and Ion Beam Processing and Fabrication}},
journal = {J. Vac. Sci. Technol. B},
volume = {26},
pages = {1197--1276},
year = {2008},
doi = {10.1116/1.2955728}
}

@book{Utke_book_2012,
editor = {Utke, I. and Moshkalev, S. and Russell, P.},
title = {{Nanofabrication Using Focused Ion and Electron Beams}},
publisher = {Oxford University Press},
address = {New York},
year = {2012}
}

@book{DeTeresa-book2020,
editor = {{De Teresa}, Jos{\'{e}} Mar{\'{i}}a},
publisher = {IOP Publishing Ltd},
title = {{Nanofabrication: Nanolithography Techniques and Their Applications}},
year = {2020},
address = {Bristol},
doi={10.1088/978-0-7503-2608-7}
}

@article{Huth_2012_BJN.3.597,
author = {Huth, M. and Porrati, F. and Schwalb, C. and Winhold, M. and Sachser, R. and Dukic, M. and Adams, J. and Fantner, G.},
title = {{Focused Electron Beam Induced Deposition: A Perspective}},
journal = {Beilstein J. Nanotechnol.},
volume = {3}, 
pages = {597--619},
year = {2012},
doi={10.3762/bjnano.3.70}
}

@article{Utke2022_CoordChemRev,
author = {Utke, I. and Swiderek, P. and H\"oflich, K. and Madajska, K. and Jurczyk, J. and Martinović, P. and Szymańska, I. B.},
title = {{Coordination and Organometallic Precursors of Group 10 and 11: Focused Electron Beam Induced Deposition of Metals and Insight Gained from Chemical Vapour Deposition, Atomic Layer Deposition, and Fundamental Surface and Gas Phase Studies}},
journal = {Coord. Chem. Rev.},
volume = {458},
pages = {213851},
year = {2022},
doi={10.1016/j.ccr.2021.213851}
}

@article{Thorman2015,
author = {Thorman, R. M. and {Ragesh Kumar}, T. P. and  Fairbrother, D. H. and Ing{\'{o}}lfsson, O.},
title = {{The Role of Low-Energy Electrons in Focused Electron Beam Induced Deposition: Four Case Studies of Representative Precursors}},
journal = {Beilstein J. Nanotechnol.},
volume = {6},
pages = {1904--1926},
year = {2015},
doi={10.3762/bjnano.6.194}
}

@article{Huth_2021_JAP_review,
author = {Huth, M. and Porrati, F. and Barth, S.},
title = {{Living Up to Its Potential -- Direct-Write Nanofabrication with Focused Electron Beams}},
journal = {J. Appl. Phys.},
volume = {130},
pages = {170901},
year = {2021},
doi={10.1063/5.0064764}
}

@article{Winkler_2019_JAP_review, 
author = {Winkler, R. and Fowlkes, J. D. and Rack, P. D. and Plank, H.},
title = {{3D Nanoprinting via Focused Electron Beams}},
journal = {J. Appl. Phys.},
volume = {125},
pages = {210901},
year = {2019},
doi = {10.1063/1.5092372}
}

@article{Barth_2025_AdvFunctMater_review,
author = {Jochmann, N. P. and Salvador-Porroche, A. and Barth, S.},
title = {{Beyond the Beam: Exploring Charged Particle Nanoprinting}},
journal = {Adv. Funct. Mater.},
volume = {35},
pages = {e07465},
year = {2025},
doi={10.1002/adfm.202507465}
}

@article{Reisecker_FEBID_review_2024,
author = {Reisecker, V. and Winkler, R. and Plank, H.},
title = {{A Review on Direct-Write Nanoprinting of Functional 3D Structures with Focused Electron Beams}},
journal = {Adv. Funct. Mater.},
volume = {34},
pages = {2407567},
year = {2024},
doi = {https://doi.org/10.1002/adfm.202407567}
}

@article{Barth2020_JMaterChemC,
author = {Barth, S. and Huth, M. and Jungwirth, F.},
title = {{Precursors for Direct-Write Nanofabrication with Electrons}},
journal = {J. Mater. Chem. C},
volume = {8},
pages = {15884--15919},
year = {2020},
doi = {10.1039/D0TC03689G}
}

@article{Sushko2016_IDMD,
author    = {Sushko, G. B. and Solov’yov, I. A. and Solov’yov, A. V.},
title     = {{Molecular Dynamics for Irradiation Driven Chemistry: Application to the FEBID Process}},
journal   = {Eur. Phys. J. D},
year      = {2016},
volume    = {70},
pages     = {217},
doi       = {10.1140/epjd/e2016-70283-5}
}

@article{Sushko2016_rCHARMM,
author    = {Sushko, G. B. and Solov’yov, I. A. and Verkhovtsev, A. V. and Volkov, S. N. and Solov’yov, A. V.},
title     = {{Studying Chemical Reactions in Biological Systems with MBN Explorer: Implementation of Molecular Mechanics with Dynamical Topology}},
journal   = {Eur. Phys. J. D},
year      = {2016},
volume    = {70},
pages     = {12},
doi       = {10.1140/epjd/e2015-60424-9}
}

@article{DeVera2020,
author = {de Vera, P. and Azzolini, M. and Sushko, G. and Abril, I. and Garcia-Molina, R. and Dapor, M. and Solov'yov, I. A. and Solov'yov, A. V.},
title = {{Multiscale Simulation of the Focused Electron Beam Induced Deposition Process}},
journal = {Sci. Rep.},
volume = {10},
pages = {20827},
year = {2020},
doi = {10.1038/s41598-020-77120-z}
}

@article{Prosvetov2021_BJN,
author = {Prosvetov, A. and Verkhovtsev, A. V. and Sushko, G. and Solov'yov, A. V.},
title = {{Irradiation-Driven Molecular Dynamics Simulation of the FEBID Process for Pt(PF$_3$)$_4$}},
journal = {Beilstein J. Nanotechnol.},
volume = {12},
pages = {1151--1172},
year = {2021},
doi = {10.3762/bjnano.12.86}
}

@article{Prosvetov2022_PCCP,
author = {Prosvetov, A. and Verkhovtsev, A. V. and Sushko, G. and Solov'yov, A. V.},
title = {{Atomistic Simulation of the FEBID-Driven Growth of Iron-Based Nanostructures}},
journal = {Phys. Chem. Chem. Phys.},
volume = {24},
pages = {10807--10819},
year = {2022},
doi = {10.1039/D2CP00809B}
}

@article{Prosvetov2023_EPJD,
author = {Prosvetov, A. and Verkhovtsev, A. V. and Sushko, G. and Solov'yov, A. V.},
title = {{Atomistic Modeling of Thermal Effects in Focused Electron Beam-Induced Deposition of Me$_2$Au(tfac)}},
journal = {Eur. Phys. J. D},
volume = {77},
pages = {15},
year = {2023},
doi = {10.1140/epjd/s10053-023-00598-5}
}

@MISC{SD_FEBID_arXiv,
author = {Solov'yov, I. A. and Prosvetov, A. and Sushko, G. and Solov'yov, A. V.},
title = {{Stochastic Dynamics Simulation of the Focused Electron Beam Induced Deposition Process}},  
url = {https://arxiv.org/abs/2506.18163},
year = {(accessed 2026-03-01)}
}

@article{DeVera2019,
author = {{de Vera}, P. and Verkhovtsev, A. and Sushko, G. and Solov'yov, A. V.},
title = {{Reactive Molecular Dynamics Simulations of Organometallic Compound W(CO)$_6$ Fragmentation}},
journal = {Eur. Phys. J. D},
volume = {73},
pages = {215},
year = {2019},
doi={10.1140/epjd/e2019-100232-9}
}

@article{Andreides_2023_JPCA_FeCO5,
author = {Andreides, B. and Verkhovtsev, A. V. and Fedor, J. and Solov’yov, A. V.},
title = {{Role of the Molecular Environment in Quenching the Irradiation-Driven Fragmentation of Fe(CO)$_5$: A Reactive Molecular Dynamics Study}},
journal = {J. Phys. Chem. A},
volume = {127},
pages = {3757--3767},
year = {2023},
doi = {10.1021/acs.jpca.2c08756}
}

@article{Lyshchuk_2025_JPCA_MeCpPtMe3,
author = {Lyshchuk, H. and Verkhovtsev, A. V. and Kočišek, J. and Fedor, J. and Solov’yov, A. V.},
title = {{Release of Neutrals in Electron-Induced Ligand Separation from MeCpPtMe$_3$: Theory Meets Experiment}},
journal = {J. Phys. Chem. A},
volume = {129},
pages = {2016--2023},
year = {2025},
doi = {10.1021/acs.jpca.4c08259}
}

@article{Jureddy_PCCP_2025,
author = {Jureddy, C. S. and  Jurczyk, J. and Mackosz, K. and Lyschuk, H. and  Kočišek, J. and Weber, P. and Ernst, M. and Verkhovtsev, A. V. and Solov'yov, A. V. and Fedor, J. and Utke, I.},
title = {{Desorption of Fragments upon Electron Impact on Adsorbates: Implications for Electron Beam Induced Deposition}},
journal = {Phys. Chem. Chem. Phys.},
volume = {27},
issue = {42},
pages = {22734--22745},
year = {2025},
doi = {10.1039/D5CP02552D}
}

@Article{FEBIMS,
AUTHOR = {Jurczyk, J. and Pillatsch, L. and Berger, L. and Priebe, A. and Madajska, K. and Kapusta, C. and Szymańska, I. B. and Michler, J. and Utke, I.},
TITLE = {{In Situ Time-of-Flight Mass Spectrometry of Ionic Fragments Induced by Focused Electron Beam Irradiation: Investigation of Electron Driven Surface Chemistry inside an SEM under High Vacuum}},
JOURNAL = {Nanomaterials},
VOLUME = {12},
YEAR = {2022},
pages = {2710},
DOI = {10.3390/nano12152710}
}

@article{Solovyov_2012_JCC_MBNExplorer,
  author = {Solov'yov, I. A. and Yakubovich, A. V. and Nikolaev, P. V. and Volkovets, I. and Solov'yov, A. V.},
  title = {{MesoBioNano Explorer -- A Universal Program for Multiscale Computer Simulations of Complex Molecular Structure and Dynamics}},
  journal = {J. Comput. Chem.},
  volume = {33},
  pages = {2412--2439},
  year = {2012},
  doi = {10.1002/jcc.23086}
}

@book{MBNbook_Springer_2017,
author = {Solov'yov, I. A. and Korol, A. V. and Solov'yov, A. V.},
title = {{Multiscale Modeling of Complex Molecular Structure and Dynamics with MBN Explorer}},
publisher = {Springer International Publishing},
address = {Cham, Switzerland},
year = {2017},
doi = {10.1007/978-3-319-56087-8}
}

@article{Sushko_2019_MBNStudio,
author = {Sushko, G. B. and Solov'yov, I. A. and Solov'yov, A. V.},
title = {{Modeling MesoBioNano Systems with MBN Studio Made Easy}},
journal = {J. Mol. Graph. Model.},
volume = {88},
pages = {247--260},
year = {2019},
doi={10.1016/j.jmgm.2019.02.003}
}

@book{DySoN_book_Springer_2022,
address = {Cham, Switzerland},
editor = {Solov'yov, I. A. and Verkhovtsev, A. V. and Korol, A. V. and Solov'yov, A. V.},
publisher = {Springer International Publishing},
title = {{ Dynamics of Systems on the Nanoscale}},
year = {2022},
doi={10.1007/978-3-030-99291-0}
}

@article{Roadmap_ChemRev2024,
author    = {Solov’yov, A. V. and Verkhovtsev, A. V. and Mason, N. J. and Amos, R. A. and Bald, I. and Baldacchino, G. and Dromey, B. and Falk, M. and Fedor, J. and Gerhards, L. and Hausmann, M. and Hildenbrand, G. and Hrabovský, M. and Kadlec, S. and Ko\v{c}i\v{s}ek, J. and Lépine, F. and Ming, S. and Nisbet, A. and Ricketts, K. and Sala, L. and Schlath\"olter, T. and Wheatley, A. E. H. and Solov’yov, I. A.},
title     = {{Condensed Matter Systems Exposed to Radiation: Multiscale Theory, Simulations, and Experiment}},
journal   = {Chem. Rev.},
year      = {2024},
volume    = {124},
pages     = {8014--8129},
doi       = {10.1021/acs.chemrev.3c00902}
}

@book{MBNTutorials_50,
author = {Verkhovtsev, A. and Solov'yov, I. A. and Korol, A. V. and Sushko, G. B. and Solov'yov, A. V.},
title = {{MBN Explorer and MBN Studio Tutorials: Version 5.0}},
publisher = {MesoBioNano Science Publishing},
address = {Frankfurt am Main, Germany},
year = {2024}
}

@article{WNOROSKI-R,
author = {Wnorowski, K. and Stano, M. and Matias, C. and Denifl, S. and Barszczewska, W. and Matej\v{c}\'ik, {\v{S}}.},
title = {{Low-Energy Electron Interactions with Tungsten Hexacarbonyl -- W(CO)$_6$}},
journal = {Rapid Commun. Mass Spectrom.},
volume = {26},
pages = {2093--2098},
year = {2012},
doi = {10.1002/rcm.6324}
}

@article{WNOROWSKI-IJMS,
author = {Wnorowski, K. and Stano, M. and Barszczewska, W. and J\'owko, A. and Matej\v{c}ík, {\v{S}}.},
title = {{Electron Ionization of W(CO)$_6$: Appearance Energies}},
journal = {Int. J. Mass Spectrom.},
volume = {314},
pages = {42--48},
year = {2012},
doi = {10.1016/j.ijms.2012.02.002}
}

@article{Rosenberg_2013_PCCP.15.4002,
author = {Rosenberg, S. G. and Barclay, M. and Fairbrother, D. H},
title = {{Electron Induced Reactions of Surface Adsorbed Tungsten Hexacarbonyl (W(CO)$_6$)}},
journal = {Phys. Chem. Chem. Phys.},
volume = {15},
pages = {4002--4015},
year = {2013},
doi = {10.1039/C3CP43902J}
}

@article{MEENUPANDEY,
author = {Meenu Pandey and Bobby Antony},
title = {{Calculations of Electron Scattering Cross Sections From Tungsten Precursors used in FEBID}},
journal = {J. Electron Spectrosc. Relat. Phenom.},
volume = {271},
pages = {147430},
year = {2024},
doi = {10.1016/j.elspec.2024.147430}
}

@article{SCOPE,
  author  = {Vinodkumar, M. and Joshipura, K. N. and Limbachiya, C. G. and Antony, B. K.},
  title   = {{Electron Impact Total and Ionization Cross-sections for Some Hydrocarbon Molecules and Radicals}},
  journal = {Eur. Phys. J. D},
  volume  = {37},
  pages   = {67--74},
  year    = {2006},
  doi     = {10.1140/epjd/e2005-00257-7}
}

@article{CSP-IC,
  author  = {Vinodkumar, M. and Korot, K. and Vinodkumar, P. C.},
  title   = {{Complex Scattering Potential--Ionization Contribution (CSP-IC) Method for Calculating Total Ionization Cross Sections on Electron Impact}},
  journal = {Eur. Phys. J. D},
  volume  = {59},
  pages   = {379--387},
  year    = {2010},
  doi     = {10.1140/epjd/e2010-00140-6}
}

@book{Ullrich_TDDFT_book,
address = {Oxford},
author = {Ullrich, C. A.},
publisher = {Oxford University Press},
title = {{Time-Dependent Density-Functional Theory: Concepts and Applications}},
year = {2012},
doi={10.1093/acprof:oso/9780199563029.001.0001}
}

@book{Marx_Hutter_AIMD_2009,
address = {Cambridge},
author = {Marx, D. and Hutter, J.},
publisher = {Cambridge University Press},
title = {{Ab Initio Molecular Dynamics. Basic Theory and Advanced Methods}},
year = {2009}
}

@book{Springer_Handbook_CompChem,
address = {Cham},
editor = {Leszczynski, J. and Kaczmarek-Kedziera, A. and Puzyn, T. and Papadopoulos, M. G. and Reis, H. and Shukla, M. K.},
edition = {2},
publisher = {Springer International Publishing},
title = {{Handbook of Computational Chemistry}},
year = {2017},
doi={10.1007/978-3-319-27282-5}
}

@article{lanl2dz_ref,
author = {Hay, P. Jeffrey and Wadt, Willard R.},
title = {{Ab initio Effective Core Potentials for Molecular Calculations. Potentials for the Transition Metal Atoms Sc to Hg}},
journal = {J. Chem. Phys.},
volume = {82},
pages = {270--283},
year = {1985},
doi = {10.1063/1.448799}
}

@article{B3LYP,
author = {Becke, Axel D.},
title = {{A New Mixing of Hartree–Fock and Local density‐Functional Theories}},
journal = {J. Chem. Phys.},
volume = {98},
pages = {1372--1377},
year = {1993},
doi = {10.1063/1.464304}
}

@article{ReaxFF_Senftle_2016,
author = {T. P. Senftle and S. Hong and M. M. Islam and S. B. Kylasa and Y. Zheng and Y. K. Shin and C. Junkermeier and R. Engel-Herbert and M. J. Janik and H. M. Aktulga and T. Verstraelen and A. Grama and A. C. T. {van Duin} },
title = {{The ReaxFF Reactive Force-Field: Development, Applications and Future Directions}},
journal = {npj Comput. Mater.},
year = {2016},
volume = {2},
pages = {15011},
doi = {10.1038/npjcompumats.2015.11}
}

@article{Mayo1990,
  author = {Mayo, S. L. and Olafson, B. D. and Goddard, W. A.},
  title = {{DREIDING: A Generic Force Field for Molecular Simulations}},
  journal = {J. Phys. Chem.},
  volume = {94},
  pages = {8897--8909},
  year = {1990},
  doi={10.1021/j100389a010}
}

@article{Oishi_2021_CompCondMatter,
  author = {Oishi, Talukder Musfika Tasnim and Malakar, Prottay and Islam, Mahmudul and Islam, Md Mahbubul},
  title = {{Atomic-Scale Perspective of Mechanical Properties and Fracture Mechanisms of Graphene/WS$_2$/Graphene Heterostructure}},
  journal = {Comput. Condens. Matter},
  volume = {29},
  pages = {e00612},
  year = {2021},
  doi={10.1016/j.cocom.2021.e00612}
}

@article{Wu_2024_JPCM,
  author = {Wu, Fan and Tan, Huifeng and Palummo, Maurizia and Camilli, Luca},
  title = {{Mechanical Properties of Bilayer WS$_2$ and Graphene-WS$_2$ Hybrid Composites by Molecular Dynamics Simulations}},
  journal = {J. Phys.: Condens. Matter},
  volume = {36},
  pages = {225301},
  year = {2024},
  doi={10.1088/1361-648X/ad2886}
}

\end{document}